\documentclass[a4paper,11pt]{amsart}

\usepackage{amsmath,amssymb,amsthm,graphicx,stmaryrd}
\usepackage{bbm}

\usepackage[utf8]{inputenc}
\usepackage[T1]{fontenc}
\usepackage[swedish, english]{babel}

\usepackage{verbatim,xcolor}
\usepackage[shortlabels]{enumitem}
\usepackage{mathtools,cancel}
\usepackage{MnSymbol}
\usepackage{hyperref}
\usepackage[margin=2.5cm]{geometry}

\usepackage[font=sf]{caption}

\numberwithin{equation}{section}

\usepackage{tikz,tikz-cd,ifthen,pgfmath}
\usetikzlibrary{patterns}
\usetikzlibrary{calc}
\usetikzlibrary{arrows.meta}
\usetikzlibrary{decorations.markings}
\usetikzlibrary{decorations.pathmorphing}

\newcommand{\bb}[1]{\mathbb{#1}} 
\renewcommand{\c}[1]{\mathcal{#1}}

\renewcommand{\bf}[1]{\mathbf{#1}}
\newcommand{\bbm}[1]{\mathbbm{#1}}
\newcommand{\dist}{\mathrm{D}}

\newcommand{\PP}{\bb{P}}
\newcommand{\EE}{\bb{E}}

\newcommand{\e}[1]{\mathrm e^{#1}}

\newcommand{\Om}{\Omega}
\newcommand{\om}{\omega} 
 
\newcommand{\s}{\sigma}

\newcommand{\ssc}[1]{\scriptscriptstyle{#1}}

\newcommand{\oo}{\infty}

\newcommand{\bra}[1]{\langle#1|}
\newcommand{\ket}[1]{|#1\rangle}

\newcommand{\se}{\subseteq}
\newcommand{\ul}{\underline}

\newcommand{\ol}{\overline}

\newcommand{\floor}[1]{\lfloor #1\rfloor}

\newcommand{\NN}{\mathbb{N}}
\newcommand{\QQ}{\mathbb{Q}}
\newcommand{\RR}{\mathbb{R}}
\newcommand{\TT}{\mathbb{T}}
\newcommand{\ZZ}{\mathbb{Z}}

\newcommand{\Tr}{\mathrm{Tr}}

\newcommand{\one}{\hbox{\rm 1\kern-.27em I}}
\newcommand{\Ind}[1]{\one_{\{#1\}}}

\newcommand{\be}{\begin{equation}}
	\newcommand{\ee}{\end{equation}}
\newcommand{\bes}{\begin{equation*}}
	\newcommand{\ees}{\end{equation*}}

\renewcommand{\r}[1]{\mathrm{#1}}
\newcommand{\dd}{\r d}

\newcommand{\p}{\r{per}}

\newcommand{\cross}{\mathchoice
	{\vcenter{\hbox{\includegraphics[width=.8em]{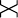}}}}
	{\vcenter{\hbox{\includegraphics[width=.8em]{cross.pdf}}}}
	{\vcenter{\hbox{\includegraphics[width=.6em]{cross.pdf}}}} 
	{\vcenter{\hbox{\includegraphics[width=.5em]{cross.pdf}}}} 
}
\newcommand{\dbar}{\mathchoice
	{\vcenter{\hbox{\includegraphics[width=.8em]{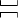}}}}
	{\vcenter{\hbox{\includegraphics[width=.8em]{dbar.pdf}}}}
	{\vcenter{\hbox{\includegraphics[width=.6em]{dbar.pdf}}}} 
	{\vcenter{\hbox{\includegraphics[width=.5em]{dbar.pdf}}}} 
}

\newcommand{\pattern}[4]{\begin{smallmatrix}#1&#2\\#3&#4\end{smallmatrix}}

\newtheoremstyle{slthm}
{}
{\baselineskip}
{\slshape}
{\parindent}
{\scshape}
{.}
{ }
{}

\theoremstyle{slthm}
\newtheorem{definition}{Definition}[section]
\newtheorem{theorem}[definition]{Theorem}
\newtheorem{proposition}[definition]{Proposition}
\newtheorem{lemma}[definition]{Lemma}

\newtheorem{remark}[definition]{Remark}

{

	\title[Exponential decay]
	{Exponential decay
		in $\c O(n)$-invariant quantum spin systems} 
	\author{J. E. Bj\"ornberg}
	\thanks{JEB: Chalmers University of Technology and University of
		Gothenburg, Sweden. jakob.bjornberg@gu.se}
	\author{K. Ryan}
	\thanks{KR: TU Wien, Austria. kieran.ryan@tuwien.ac.at}
	
	\date{\today}

\begin{document}
		
		\begin{abstract}
			We consider $\c O(n)$-invariant and reflection-positive quantum spin
			systems on the integer lattice in any dimension, and prove that
			spin-spin correlations decay exponentially fast provided $n$ is large
			enough.  This answers a question of Ueltschi, who proved that 
			for small $n$ there is instead
			long-range order  (for $d\geq3$).  
		\end{abstract}
		
		\maketitle

		\section{Introduction}
		
		The most general $\c O(n)$-invariant quantum spin system with pair-interactions has Hamiltonian
		\be\label{eq:ham-general}
		H_\Lambda=-\sum_{xy\in\c E(\Lambda)}
		\big[
		u T_{xy}+v Q_{xy}
		\big],
		\qquad\text{acting on } (\bb C^n)^{\otimes \Lambda},
		\ee
		where $u,v\in\RR$,
		$\Lambda$ is a finite graph,
		$\c E(\Lambda)$ is its set of nearest-neighbour pairs,
		and the interaction involves the operators
		\be
		T\ket{a,b}=\ket{b,a},
		\qquad
		Q=\frac1{n}\sum_{a,b=1}^n \ket{b,b}\bra{a,a},
		\qquad\text{on } (\bb C^n)^{\otimes 2}.
		\ee
		The behaviour of this model varies widely depending on the graph
		$\Lambda$, the parameters $u,v$, and the value of $n$ (related to the
		spin $S$ by $2S+1=n$). The model reduces to the spin-$\tfrac{1}{2}$
		Heisenberg XXZ model when $n=2$, and to the spin-1
		bilinear-biquadratic Heisenberg model when $n=3$. For $u,v\ge0$, the
		model has a well-known probabilistic representation as a loop model
		(related to the interchange process). This paper studies the model
		on $\ZZ^d$ with $v\ge u \ge 0$ and $n$ large, where we prove
		exponential decay of spin-spin correlations. In a separate paper
		\cite{BR-dimer}, we prove a breaking of translation invariance known
		as dimerization in the ground state in dimension 1, for $u\ge0$, $v>0$
		and $n$ large. \\

		\subsection{Main result}
		
		We set $v=1-u$, so the Hamiltonian is 
		\be\label{eq:ham}
		H_\Lambda=-\sum_{xy\in\c E(\Lambda)}
		\big[
		u T_{xy}+(1-u) Q_{xy}
		\big].
		\ee
		We let $S^{(1)},S^{(2)},S^{(3)}$ be the usual spin-operators
		(generators of $\mathfrak{su}(2)$) acting on $\bb C^n$
		and $S^{(i)}_x=S^{(i)}\otimes\one_{\Lambda\setminus\{x\}}$
		acting on $(\bb C^n)^{\otimes \Lambda}$.
		For an operator $A$ on $(\bb C^n)^{\otimes \Lambda}$ we let
		$A(t)=\e{t H_\Lambda}A\e{-tH_\Lambda}$.
		We consider the Gibbs state
		\be
		\langle A\rangle_{\Lambda,\beta}
		:=\frac{\Tr[A\e{-\beta H_\Lambda}]}{\Tr[\e{-\beta H_\Lambda}]}
		\ee
		and truncated correlations
		\be
		\langle A;B\rangle_{\Lambda,\beta}:=
		\langle AB\rangle_{\Lambda,\beta}-
		\langle A\rangle_{\Lambda,\beta}
		\langle B\rangle_{\Lambda,\beta}.
		\ee
		For $k_1,\dotsc,k_d\geq 1$
		we write $\ul k=(k_1,\dotsc,k_d)$ and set
		\phantomsection\label{not Lk}
		\be\label{eq:torus}
		\Lambda(\ul k)=\{0,\dotsc,4k_1-1\}\times
		\dotsb \times \{0,\dotsc,4k_d-1\}.
		\ee
		We regard $\Lambda(\ul k)$ as a torus of spatial dimensions $4k_1,\dots,4k_d$.
		We use the notation $\ul x=(x_1,\dotsc,x_d)$ for elements of 
		$\Lambda(\ul k)$ or $\bb Z^d$, and 
		$(\ul x,t)$ for elements of $\Lambda(\ul k)\times[0,\beta]$ 
		or $\bb Z^d\times\bb R$.
		
		\begin{theorem}[Exponential decay of correlations]\label{thm:d>1-exp-decay}
			Let $u\in[0,\tfrac{1}{2}]$ and $d\geq1$.  
			Consider the model with Hamiltonian
			\eqref{eq:ham} on the 
			torus $\Lambda(\ul k)$.
			For any $c>0$, there exist $n_d\in\NN$ and $\alpha_d>0$
			(dependent on $u,d,c$, but not on $k_1,\dots,k_d,\beta$)
			such that for all integers $n>n_d$ and all $\beta>\alpha_d/n$, 
			\be
			|\langle S^{(i)}_0;S^{(i)}_x(t)\rangle_{\Lambda(\ul k),\beta}|
			\le 
			\e{-c (\|\ul{x}\|_1 +|t|)},
			\ee
			for all $i\in\{1,2,3\}$.
		\end{theorem}

		For $i\in\{1,3\}$ the correlations are non-negative, so the
		absolute-value is only needed for $i=2$.  For all
		$i\in\{1,2,3\}$ we actually have
		$\smash{\langle S^{(i)}_0;S^{(i)}_x(t)\rangle_{\Lambda,\beta}=
			\langle S^{(i)}_0S^{(i)}_x(t)\rangle_{\Lambda,\beta}}$,
		as  follows from the symmetries in 
		\cite[Lemma 3.1]{ueltschi} (observe that Ueltschi's 
		notation for correlation
		functions differs from ours).

		Theorem \ref{thm:d>1-exp-decay}  answers a question of Ueltschi
		\cite{ueltschi}, who proved a complementary result: for
		$u\in[0,\tfrac{1}{2}]$, $d\ge3$, $n<n_1(d,u)$, and 
		$\beta>\beta_1(u,d,n)$, there is long range order (in the loop model,
		this is the presence of infinite loops in the infinite volume
		system). He then asked whether the same was true for fixed dimension
		and all spins;
		our Theorem \ref{thm:d>1-exp-decay}  answers this in the negative. 
		Betz, Klippel and 
		Nauth \cite{betz-klippel-nauth} recently improved the values of $n_1$
		and $\beta_1$.

		\begin{remark}
The upcoming Theorem \ref{thm:refpos:exp-decay-loops}
			implies exponential decay of any correlations expressible as
			probabilities of loop-connections.  For example, 
			write $E^{a,b}=\ket{b}\bra{a}$ for the elementary operator.
			Using Lemma 2.4 of \cite{BR-dimer} we can conclude that
			for any $a\neq b$, the correlation
			$\langle E_{\ul x}^{a,b}(s) E_{\ul y}^{a,b}(t)\rangle$ decays exponentially under
			the same assumptions as in Theorem \ref{thm:d>1-exp-decay}.
			
			Furthermore, in \cite[Theorem 3.5]{ueltschi} Ueltschi considers a
			closely
			related model where the operator $Q_{xy}$ in \eqref{eq:ham} is
			replaced by  another projection operator $P_{xy}$, and he proves that for
			\emph{odd values} of $n$, correlation function of the form 
			$\langle (S^{(i)}_{\ul x}(s))^2; (S^{(i)}_{\ul y}(t))^2\rangle$ in that model can
			also be written as connection probabilities.  Consequently, our
			theorem applies also to them.
		\end{remark}

For $n=3$, the model \eqref{eq:ham} is equivalent to the
		bilinear-biquadradic Heisenberg model with Hamiltonian 
		\be\label{eq:ham-bi-bq}
		H^{\text{bi-bq}}_\Lambda=-\sum_{xy\in\c E(\Lambda)}
		\big[
		u\bf{S}_x\cdot\bf{S}_y + (\bf{S}_x\cdot\bf{S}_y)^2
		\big].
		\ee\\
		(Note that the two-parameter version \eqref{eq:ham-general} gives the
		general two-parameter version of \eqref{eq:ham-bi-bq}). In this
		setting, Ueltschi's long range order is Nematic order for
		$u\in[0,\tfrac{1}{2}]$ (expected to hold for all $u\in(0,1)$) and the stronger Néel
		order for $u=0$ (proved for $d\ge5$, expected to hold for
		$d\ge3$). Lees \cite{lees-antiferro-neel} proved nematic order for $u$ small and negative in $d\ge6$ (the stronger Néel order is expected). 
		
		For $n=2$, the model is the XXZ model.
		The antiferromagnet was proved to have long range
		Néel order at low temperatures for all $n\ge3$ (spin $S\ge1$) and
		$d\ge3$, and $n=2$ ($S=\tfrac{1}{2}$) with $d$ large enough by Dyson,
		Lieb and Simon \cite{DLS}.
		Kennedy, Lieb and Shastry \cite{KLS-antiferro} completed the picture for $S=\tfrac{1}{2}$, all $d\ge3$ as well as the ground state in $d\ge2$; they proved the same for the XY model in \cite{KLS-XY}. Lees \cite{lees-nematic} works on the $n=3$ (spin $S=1$) antiferromagnet with some extra nematic interaction and proves Néel order in $d\ge3$. Ueltschi's work \cite{ueltschi} is the first known to the authors that covers the range $u\in(0,\tfrac{1}{2})$ ($\Delta\in(-1,0)$). All of these results use reflection positivity, a property only enjoyed by the model \eqref{eq:ham} for $u\in[0,\tfrac{1}{2}]$. Proving long range order at low temperatures in the range $u>\tfrac{1}{2}$ remains open; the special case $u=1$, $n=2$, the spin $S=\tfrac{1}{2}$ Heisenberg ferromagnet, is a big open question in the field.
		
		In dimension $d=2$, there is no long range order at positive temperatures due to the Mermin-Wagner theorem, originally written for the Heisenberg ferromagnet \cite{merm-wag}. This has been extended by many authors; in particular Fröhlich and Pfister showed that all KMS states are rotationally invariant \cite{fro-pfist}. Koma and Tasaki \cite{koma-tasaki} proved that correlations in many 2D quantum spin systems must decay at least polynomially fast, extending an argument of McBryan and Spencer \cite{mcbryan-spencer} for classical systems. Benassi, Fröhlich and Ueltschi \cite{BFU} extended this to many models, in particular the model \eqref{eq:ham} for all $n\ge2$, $u\in[0,1]$. Our result extends theirs to exponential decay for $n$ large, $u\in[0,\tfrac{1}{2}]$. For $n=2$, a BKT transition is expected for $u\in(0,1)$ ($\Delta\in(-1,1)$); proving a lower bound on correlations at low temperatures remains open. The points $u=0,1$ ($\Delta=-1,1$) retain $SU(2)$ symmetry and should have exponentially decaying correlations at all positive temperature. One could ask what the minimal $n$ is where one has exponential decay for all $u\in[0,1]$. Since the $n=3$ model \eqref{eq:ham-bi-bq} is $SU(2)$ invariant, perhaps the minimum is $n=3$. This would align with thinking about the loop model as a 2D loop-$O(n)$ model, which should behave like a classical spin $O(n)$ model, ie. in line with Polyakov's Conjecture.
		
		In dimension $d=1$, we have a stronger result
		on exponential decay in our separate paper \cite{BR-dimer}:
		there it is proved that \emph{all} truncated correlations decay
		exponentially, for all $u\in[0,1)$ (provided $n$ is large enough).
		The restriction $u\in[0,\tfrac{1}{2}]$ in the present paper
		comes from relying on
		reflection-positivity, a feature of the model for this range of $u$
		only \cite{ueltschi}.
		
		The proof of 
		Theorem \ref{thm:d>1-exp-decay} is based on the method of chessboard estimates,
		facilitated by the loop model's reflection positivity, and arguments 
		inspired by Chayes, Pryadko and Shtengel \cite{chayes-etal}. See
		Biskup \cite{biskup-RP} for an overview of reflection positivity for
		classical systems, and Björnberg and Ueltschi \cite{BU-RP} for quantum
		systems.\\

		In Section \ref{section:prob-rep} we define the probabilistic
		representation of the model, and state our main probabilistic result, Theorem \ref{thm:refpos:exp-decay-loops}. 
		We then show how this theorems implies Theorem
		\ref{thm:d>1-exp-decay}. 
		In
		Section \ref{sec high d}, we prove Theorem \ref{thm:refpos:exp-decay-loops}.

		\subsection*{Acknowledgements}
		
		It is a pleasure to thank
		Ron Peled, Daniel Ueltschi and Bruno Nachtergaele for
                very useful discussions about this project. 
		The research of JEB was supported by 
		Vetenskapsr{\char229}det, 
		grant 2023-05002, 
		and by \emph{Ruth och Nils Erik Stenbäcks stiftelse}.  He gratefully
		acknowledges hospitality at TU Wien and at Aalto University.
		KR was supported by the FWF Standalone grants ``Spins, Loops and
		Fields'' P 36298 and ``Order-disorder phase transition in 2D lattice
		models'' P 34713,  the FWF SFB Discrete Random Structures F 1002, and
		the Academy of Finland Centre of Excellence Programme grant number
		346315 ``Finnish centre of excellence in Randomness and STructures''
		(FiRST).  He gratefully
		acknowledges hospitality at Chalmers / University of Gothenburg.

		\section{Probabilistic framework}\label{section:prob-rep}

		Our proofs rely on a well-known probabilistic representation 
		\cite{toth,an,ueltschi} where the quantum
		system is expressed in terms of a process of random loops.
		In Section \ref{sec high d}
		we work exclusively in this probabilistic framework.  The purpose of
		the present section is to describe the random loop model, to state our
		main result for the loop model, and explain how Theorem 
		\ref{thm:d>1-exp-decay} follows from the result on the loop model.

		\subsection{Loop model}

		Recall that the quantum model is defined on a finite graph
		$\Lambda=(\c V(\Lambda)=\c E(\Lambda))$.  
		Let $\beta>0$ and $u\in[0,1)$.  Let 
		$\PP_1$ denote the law of the superposition 
		of two independent Poisson
		point processes on $\c E(\Lambda)\times[0,\beta]$, 
		the first of  intensity $u$ and 
		whose points we denote by $\cross$ (called a
		\emph{cross}), and the second of  intensity $1-u$ and 
		whose
		points we denote by $\dbar$ (called a \emph{double bar}). 
		We write $\om$ for a configuration of this
		process, and the set of such configurations is
		$\Om=\Om_{\Lambda,\beta}=(\cup_{k\ge0}\c W_k)^2$, where $\c W_k$ is
		the set of subsets of
		$\c E(\Lambda)\times[0,\beta]$ of cardinality
		$k$.   A point of  $\omega$ is called a \emph{link}.

\begin{figure}[th]
			\begin{center}
				\includegraphics[scale=.85]{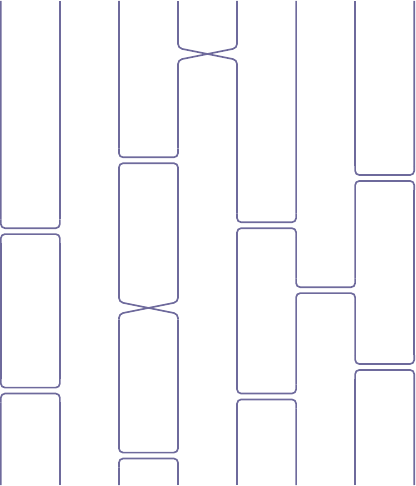}%
			\end{center}
			\caption{Example of a 
				configuration $\omega$ in dimension $d=1$.
				The links ($\cross$ and
				$\dbar$) create loops, which wrap around in the
				vertical direction.
			}
			\label{fig loops}
		\end{figure}

		Each configuration $\om$ gives a set of loops, best understood by
		looking at Figure \ref{fig loops} and defined formally as follows. 
		First, we identify the endpoints $0$ and $\beta$ of $[0,\beta]$ so
		that it forms a circle;  in particular an interval
		$(a,b]\subset[0,\beta]$ with $a>b$ is defined as
		$(a,\beta]\cup(0,b]$.
		Let $\om\in\Om$ and consider the set $\c I(\om)$ of maximal
		intervals $\{x\}\times(a,b]$, $x\in \c V(\Lambda)$ and
		$(a,b]\se[0,\beta]$, which are not adjacent to a link; that is,
		intervals $\{x\}\times(a,b]$ such that there is 
		no link at $(\{x,x'\},t)$ for all $t\in(a,b)$ and all
		$x'\sim x$.  
		
		Incident to a link of $\om$ at, say, $(\{x,x'\},t)$, there are four
		intervals in $\c I$: 
		$\{x\}\times(t,b_x]$, $\{x\}\times(a_x,t]$, 
		$\{x'\}\times(t,b_{x'}]$, and $\{x'\}\times(a_{x'},t]$.  We say that:
		\begin{itemize}[leftmargin=*]
			\item if the link of $\om$ at $(\{x,x'\},t)$ is a cross $\cross$, then
			$\{x\}\times(t,b_x]$ and $\{x'\}\times(a_{x'},t]$ are connected, 
			and $\{x\}\times(a_x,t]$ and $\{x'\}\times(t,b_{x'}]$ are connected;
			\item if the link of $\om$ at $(\{x,x'\},t)$ is a double-bar $\dbar$, then
			$\{x\}\times(t,b_x]$ and $\{x'\}\times(t,b_{x'}]$ are connected, 
			and $\{x\}\times(a_x,t]$ and $\{x'\}\times(a_{x'},t]$ are connected.
		\end{itemize}
		The loops of $\om$ are then the connected components of $\c I$ under
		this connectivity relation.  
		
		Write $\ell(\om)$ for the number of loops of $\om$.
		The loop measure
		we study is the Poisson point process $\PP_1$ re-weighed by
		$n^{\ell(\om)}$, 
		and we denote it by $\PP_{\Lambda,\beta,n,u}$: 
		\be\label{eq measure periodic}
		\PP_{\Lambda,\beta,n,u}[A]
		=\frac{1}{Z}\int \text{d} \PP_1(\om) \; 
		n^{\ell(\om)} \mathbbm{1}_A(\om),
		\ee
		where 
		$Z=Z_{\Lambda,\beta,n,u}=\int \text{d} \PP_1(\om) \; n^{\ell(\om)}$. 
		To abbreviate we sometimes write simply $\PP_n$ for 
		$\PP_{\Lambda,\beta,n,u}$;
		note that when $n=1$ we recover $\PP_1$.
		(The measure $\PP_n$ is well-defined also for
		non-integer $n$, although connection to the quantum system
		\eqref{eq:ham} requires $n$ to be an integer at least 2.
		In this article we only consider integer-valued $n$.)
		
		\subsection{Main result for the loop model}
		
		We write $(\ul{x},s) \leftrightarrow (\ul{y},t)$ for the event that
		$(\ul{x},s)$ and $(\ul{y},t)$ belong to the same loop.
		The following is the probabilistic counterpart to
		Theorem \ref{thm:d>1-exp-decay}. 
		Recall the torus $\Lambda(\ul k)$
		from \eqref{eq:torus}.

		\begin{theorem}[Exponential decay in the loop model]
			\label{thm:refpos:exp-decay-loops}
			Let $d\geq1$ and $u\in[0,\tfrac{1}{2}]$
			and consider the loop model on
			$\Lambda(\ul k)\times[0,\beta]$.  
			For any $c>0$, there exist $n_d\in\NN$ and $\alpha_d>0$
			(dependent on $u,d,c$, but not on $k_1,\dots,k_d,\beta$)
			such that for all integers $n>n_d$ and all $\beta>\alpha_d/n$, 
			\be
			\PP_n[(\ul 0,0) \leftrightarrow (\ul{x},t)]
			\le 
			\exp[-c (\|\ul{x}\|_1+|t|)].
			\ee
		\end{theorem}
		
		The basic reason for this result is that when $n$ is large, the system
		favours many loops (due to the weight factor 
		$n^{\ell(\om)}$), hence small loops.

		
		The loop model allows the expectation of local observables of the
		quantum 
		system to be written using probabilities of events in the loop
		model.
		We use the following specific examples of this,
		proved in \cite[Theorem 3.3]{ueltschi}.
		Theorem \ref{thm:d>1-exp-decay} follows immediately from 
		Lemma \ref{lem:corrs} and
		Theorem	\ref{thm:refpos:exp-decay-loops}.
		
		\begin{lemma}\label{lem:corrs}
			For all integers $n\geq2$ and $u\in[0,1]$,
			\be
			\langle S^{(1)}_{\ul x}(s); S^{(1)}_{\ul y}(t)\rangle
			=\langle S^{(3)}_{\ul x}(s); S^{(3)}_{\ul y}(t)\rangle
			=\tfrac{n^2-1}{12} \PP_n((\ul x,s)\leftrightarrow(\ul y,t))
			\ee
			and
			\be
			|\langle S^{(2)}_{\ul x}(s); S^{(2)}_{\ul y}(t)\rangle|\leq
			\tfrac{n^2-1}{12} \PP_n((\ul x,s)\leftrightarrow(\ul y,t)).
			\ee
		\end{lemma}

		We note that Theorem \ref{thm:refpos:exp-decay-loops}
		proves exponential decay only for $\beta>\alpha_d/n$, 
		where $\alpha_d$ is some large constant depending on $d$. 
		Stochastic domination by a Poisson process of rate $n$ 
		combined with  a standard percolation
		argument gives exponential decay for all $\beta<\alpha'_d/n$, where
		$\alpha'_d$ is a \emph{small} constant depending on $d$.
		We certainly expect the exponential decay to hold for all $\beta>0$,
		but the gap
		$\beta\in[\alpha'/n,\alpha/n]$ remains. The restriction
		$\beta>\alpha_d/n$ in our theorem is an artefact of the proof, in
		particular the change in boundary conditions in
		\eqref{eq:crowded-boundary-conditions} in the proof of Lemma
		\ref{refpos:lem:bound-bad-events}.  Avoiding this
		technique would yield a proof for all $\beta>0$, but it is unclear how
		to do so.

		\section{Proof of exponential decay in the loop model}
		\label{sec high d}

		The proof of Theorem \ref{thm:refpos:exp-decay-loops}
		uses a Peierls argument, facilitated by 
		\emph{chessboard estimates}, a feature of reflection positivity. 
		The event $(\ul{0},0)
		\leftrightarrow (\ul{x},t)$ implies the existence of a path $\Gamma$
		in a discrete torus $\TT$ (defined below), such that on a positive
		fraction of the blocks in $\Gamma$, a `bad' event $B$ (also defined
		below) occurs.  The probability of the bad event $B$ is shown to be
		small using the chessboard estimate, essentially because it
		implies a `local deficit' of loops.  Another application of the
		chessboard estimate shows that 
		bad paths $\Gamma$ are exponentially unlikely in their length,
		which allows to bound the probability of their existence
		by summing over all possibilities.
		
		We arrange the proof as follows.
		The two key steps, i.e.\ showing that $\Gamma$ exists and  bounding
		the probability of bad events, are formulated as 
		Lemmas \ref{refpos:lem:SAW-exists} and
		\ref{refpos:lem:bound-bad-events}.
		The chessboard estimate
		itself is stated as  Proposition \ref{refpos:prop:chessboard}. 
		These appear in Section
		\ref{subsec:refpos:part-1}, where we also
		detail the Peierls argument.  In Sections \ref{subsec:refpos:part-2} and
		\ref{subsec:refpos:part-3} we prove Lemmas \ref{refpos:lem:SAW-exists} and
		\ref{refpos:lem:bound-bad-events}, respectively.  
		Finally, in Section \ref{subsec:refpos:chessboard} we prove reflection
		positivity and the chessboard estimate.

		\subsection{Auxiliary results and Peierls argument}
		\label{subsec:refpos:part-1}
		
		We think of $\Lambda(\ul k)\times[0,\beta]$ as a subset
		of the `continuous' torus 
		$\c T=[0,4k_1]\times\cdots\times[0,4k_d]\times[0,\beta]$.
		For nearest neighbours
		$\ul x,\ul x'\in \Lambda(\ul k)$ 
		the corresponding edge $e=\{\ul x,\ul x'\}$
		is identified with the midpoint $\frac12(\ul x+\ul x')$.
		The random configuration $\omega$ then consists of points on
		intervals of the form  $\tfrac12(\ul x+\ul x')\times[0,\beta]$, with labels
		$\dbar$ or $\cross$.
		
\begin{figure}[th]
			\begin{center}
				\includegraphics[scale=.85]{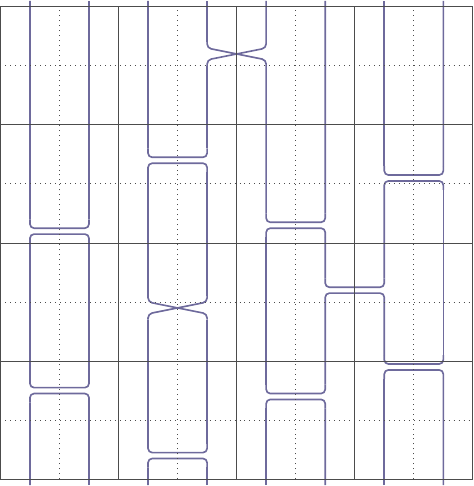}%
			\end{center}
			\caption{The same configuration $\om$ as in
                          Figure \ref{fig loops} but with big cubes
                          (solid outlines) and small cubes (dotted
                          outlines) drawn.
			}
			\label{fig cubes}
		\end{figure}

		Let $R_0>0$ and assume that $\beta>2R_0/n$.  Let $R$ be the smallest
		value $R>R_0$ such that $2k_{d+1}:=\beta n/R$ is an even integer. 
		We divide the torus 
		$[0,4k_1]\times\cdots\times[0,4k_d]\times[0,\beta]$
		into two discrete tori, $\bb T$ and $\bbm t$.
		These are defined in terms of what we call \emph{big cubes} and
		\emph{small cubes},
		collectively \emph{cubes}.   For an illustration of these, see
		Figure \ref{fig cubes}.
		
		By \emph{big cubes} we mean the hypercubes
		\be
		S_{\ul{i},j} :=
		[i_1-\tfrac{1}{2},i_1+\tfrac{3}{2}]\times
		\cdots
		\times[i_d-\tfrac{1}{2},i_d+\tfrac{3}{2}]
		\times[j\tfrac{R}{n},(j+1)\tfrac{R}{n}]\se\bb R^{d+1},
		\ee
		where $0\le i_r\le 4k_r-1$ are all even, and $0\le j \le 2k_{d+1}-1$.
		Thus the cubes  $S_{\ul i,j}$ have  spatial width 2 and height $R/n$
		and there are $K_{d+1}$ of them, where 
		$K_{d+1}=\prod_{r=1}^{d+1} 2k_r$.    
		We regard two cubes $S_{\ul i,j}$  and $S_{\ul i',j'}$ as \emph{adjacent} if
		either $j,j'$ differ by 1 and $\ul i'=\ul i$, 
		or $j'=j$ and exactly  one of the $i_r,i'_r$ differ by
		2, in both cases counting periodically.  
		We let $\TT$ be the discrete torus made up of the $K_{d+1}$ big cubes
		$S_{\ul i,j}$.
		We also need to consider the $2^{d+1}$ translates of $\bb T$
		given by translating (periodically) by a vector 
		$(\ul{i},j\tfrac{R}{2n})$, with
		$i_1,\dotsc,i_d,j\in\{0,1\}$.   We label these translates as simply
		$\TT_1,\dots,\TT_{2^{d+1}}$.
		
		Secondly, we will need a discrete torus $\bbm t$ corresponding to a
		finer partition into \emph{small} cubes.  
		By \emph{small cubes} we mean the
		hypercubes 
		\be
		s_{\ul{i},j}:=
		[i_1-\tfrac{1}{2},i_1+\tfrac{1}{2}]\times
		\cdots
		\times[i_d-\tfrac{1}{2},i_d+\tfrac{1}{2}]
		\times[j\tfrac{R}{2n},(j+1)\tfrac{R}{2n}]\se\bb R^{d+1},
		\ee 
		where $0\le i_r\le 4k_r-1$, not necessarily even, 
		and $0\le j \le 4k_{d+1}-1$.  
		Adjacency of small cubes is defined in the natural way.
		Each big cube
		$S_{\ul{i},j}$ of $\TT$ is the union of exactly $2^{d+1}$ small cubes of
		$\bbm t$. In our figures
		(e.g.\ Figure \ref{fig cubes})
		we have $d=1$, so a cube of $\TT$ is drawn
		as 4 cubes of $\bbm t$ joined together. 
		Since cubes (both big and small) are defined as \emph{closed} sets,
		links on the boundary of a cube are considered to lie in that cube.
		
		By slight abuse of terminology,
		we consider a vertex $\ul x\in\Lambda(\ul k)$ to lie in a
		cube  $S_{\ul i,j}$ or $s_{\ul i,j}$ if there is a
		$t\in[0,\beta]$ such that $(\ul x,t)\in\bb R^{d+1}$ 
		lies in the cube.  Similarly, an edge (nearest-neighbour pair)
		of $\Lambda(\ul k)$ is considered  to
		lie in a cube if either of its endpoint vertices lies in the cube. Note that
		each cube has several edges which also belong to neighbouring
		cubes. 
		
		Write $\pi$ for the set of hyperplanes which are each
		orthogonal to one of the coordinate
		axes of $\bb R^{d+1}$ and
		which exactly pass through the boundaries of cubes in $\TT$. For
		$q\in\TT$, write $\theta_q$ for the composition of reflections in
		hyperplanes in $\pi$ which maps $S_{\ul{0},0}$ to $S_q$ (note that
		the map this produces is independent of the choice of reflections to
		compose).  
		
		We say that an event $A$ \emph{occurs in} a set $U\se\bb R^{d+1}$ if
		it is dependent only on the configuration $\om$ restricted to $U$. For
		an event $A$ occurring in $S_{\ul{0},0}$ and $q\in\TT$, write
		$\theta_q A=\{\theta_q\om \ : \ \om\in A\}$.
		Note that $\theta_qA$ occurs in $S_q$.  We also say
		that $A$ occurs in the cube $q\in\TT$ if $\theta_qA$ occurs.
		We define the \emph{distributed event}
		\be
		\dist A:=\bigcap_{q\in\TT}\theta_qA.
		\ee 
		
		We say that an event $A$ is
		\emph{independent of link type in} $U\se\bb R^{d+1}$
		if for all $\om\in A$, and for all $\om'$
		obtainable from $\om$ by changing the type of any number of links of
		$\om\cap U$ (i.e.\ double bar $\dbar$ to cross $\cross$ 
		or vice-versa), we have $\om'\in A$
		(thus, for links in $U$, the event depends only on their
		number and locations, not on their types).

		\begin{proposition}[Chessboard estimate]\label{refpos:prop:chessboard}
			Let $A_1,\dots,A_m$ be events occurring in $S_{\ul{0},0}$ that are
			independent of link type on the boundary of $S_{\ul 0,0}$.  Then  for any
			distinct $q_1,\dots,q_m\in\TT$,  
			\bes
			\PP_n\Big[
			\bigcap_{i=1}^m \theta_{q_i} A_i
			\Big]
			\le
			\prod_{i=1}^m
			\PP_n\big[ \dist A_i 
			\big]^{1/K_{d+1}}.
			\ees
		\end{proposition}
		
		The condition that the events are independent of link type on the
		boundary of $S_{\ul{0},0}$ is necessary, and comes from the fact that
		the measure $\PP_n$ itself is not reflection positive, but instead
		another measure to which $\PP_n$ is coupled. See Section
		\ref{subsec:refpos:chessboard} for details.  Note that the events
		$A_i$ have  no restrictions on the configuration on
		edges in the interior of $S_{\ul{0},0}$.

\begin{figure}[th]
			\begin{center}
					\includegraphics{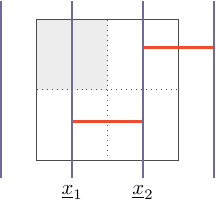}%
					\hspace{3cm}
					\includegraphics{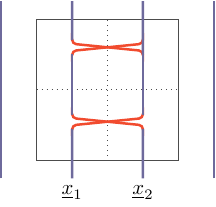}
			\end{center}
			\caption{Illustration of the  three  bad events $C$, $E$, $T$.
				The big cube $S_{\ul 0,0}$ is drawn with a solid outline, its four
				constituent small cubes are indicated with dotted lines.
				Left: a configuration in $C\cap E$.
				The two edges with common endpoint $\ul x_2$ both have links in
				$S_{\ul 0,0}$, 
				as required for $C$, and the top left small cube 
				containing $\ul x_1$ (shaded grey) has no links, as required for $E$.
				Right: the transposition event $T$, where at least one
				cross $\cross$ appears in the interior of $S_{\ul 0,0}$.
			}
			\label{bad-events}
		\end{figure}

		We now define the following three `bad' events (see Figure \ref{bad-events}):
		\begin{itemize}[leftmargin=*]
			\item \emph{Crowded event} $C$: 
			there is a pair of adjacent edges $e_1\neq e_2$ (i.e.\ sharing exactly
			one endpoint)  in $S_{\ul{0},0}$  which both have a link in $S_{\ul{0},0}$;
			\item \emph{Empty event} $E$:  one of the small cubes
			$s_{\ul{i},j}$ of $\bbm t$ contained in $S_{\ul{0},0}$ contains no
			links in $S_{\ul{0},0}$; 
			\item \emph{Transposition event} $T$: there is a cross
			$\cross$ lying on some edge $e$ in the interior of
			$S_{\ul{0},0}$. 
		\end{itemize}
		Note that all three of these events are independent of link type on
		the boundary of $S_{\ul{0},0}$, so the chessboard
		estimate, Proposition \ref{refpos:prop:chessboard}, applies to them.
		We define $B:=C\cup E\cup T$.
		Intuitively, a configuration not in $B$ has double bars closely
		stacked on disjoint sets of edges.

		\begin{lemma}\label{refpos:lem:SAW-exists}
			On the event $(\ul{0},0) \leftrightarrow (\ul{x},t)$
			there is some $i\in\{1,\dotsc,2^{d+1}\}$ such that $\TT_i$
			contains a path $\Gamma$ where
			in a fraction at least 
			$\varphi=(2(d+1)^2+1)^{-1}$ of the cubes
			of $\Gamma$, the 
			event $B$ occurs. 
		\end{lemma}

		\begin{lemma}\label{refpos:lem:bound-bad-events}
			For any $R>0$
			there exists $C_1=C_1(d,R)>0$ such that 
			for $\beta>20 \cdot 2^{d+1}R/n$ and $n$ large enough (given $R$ and
			$d$) 
			\be
			\begin{split}
				\PP_n[\dist E]^{1/K_{d+1}}&\le
				2^{d+2} \e{-(1-u)R/4}\\
				\PP_n[\dist T]^{1/K_{d+1}}
				&\le \e{-d2^d R} + (C_1(1-u)n)^{-1/5},\\
				\PP_n[\dist C]^{1/K_{d+1}}&\leq 
				2^d{2d\choose 2}
				\big(\e{-d2^d R}+   (C_1(1-u)n)^{-1/5}\big)
			\end{split}
			\ee
		\end{lemma}
		
		\noindent
		With these lemmas in hand we can  run the Peierls argument to prove
		Theorem \ref{thm:refpos:exp-decay-loops}.

		\begin{proof}[Proof of Theorem \ref{thm:refpos:exp-decay-loops}]
			We write $L=\|(\ul{x},t)\|_1$ and we say that a cube $q\in\bb T$
			is \emph{bad} if $B$ occurs in $q$.  
			To save space we write $O$ for $(\ul 0,0)$ and $X$ for $(\ul x,t)$.
			Note that the maximum $\|\cdot\|_1$-distance between points in
			neighbouring cubes of $\bb T^\ast$ is $\Delta:=3d+R/n$
			(for example $(\ul 0,0)\in S_{\ul 0,0}$ to 
			$(\ul 3,R/n)\in S_{\ul 2,0}$) so a path $\Gamma$ in $\bb T$ from
			$O$ to $X$ consists of at least $L/\Delta$ cubes.
			Applying a union bound over the possible tori 
			$\bb T_1,\dotsc,\bb T_{2^{d+1}}$ 
			in Lemma \ref{refpos:lem:SAW-exists}
			and then summing over the possibilities for the path 
			$\Gamma$ and the subset of
			bad cubes in $\Gamma$ we get
			\be\label{refpos:eq:peierls:1}
			\begin{split}
				\PP_n[O \leftrightarrow X] 
				&\le 
				2^{d+1}  \PP_n[\exists \ \Gamma:O \to X \ 
				\text{in } \TT, \ \ge\varphi \mathrm{\ of \ } \Gamma \ \mathrm{bad}]\\ 
				&\le 2^{d+1}
				\sum_{N\ge L/\Delta}
				\sum_{\substack{\Gamma:O\to X \\|\Gamma|=N}} 
				\sum_{\{q_1,..,q_{\floor{\varphi N}}\}\se\Gamma}
				\PP_n\Big[\bigcap_{i=1}^{\floor{\varphi N}}\theta_{q_i}B\Big].
			\end{split}\ee
			Applying the chessboard estimate, Proposition
			\ref{refpos:prop:chessboard}, we bound
			\be\label{refpos:eq:peierls:2}
			\PP_n\Big[\bigcap_{i=1}^{\floor{\varphi N}}\theta_{q_i}B\Big]
			\leq
			\prod_{i=1}^{\floor{\varphi N}} 
			\PP_n\Big[\bigcap_{q\in\TT}\theta_{q}B\Big]^{1/K_{d+1}}
			=\prod_{i=1}^{\floor{\varphi N}} \PP_n\big[\dist B\big]^{1/K_{d+1}}
			\ee
			where the right-hand-side no longer depends on the choice of subset 
			$\{q_1,..,q_{\floor{\varphi N}}\}$.  Since $B=C\cup E\cup T$ is a union of
			events, we can apply subadditivity
			of the chessboard estimate \cite[Lemma~5.9]{biskup-RP}:
			\be
			\PP_n\big[\dist B\big]^{1/K_{d+1}}\leq 
			\sum_{A\in\{C,E,T\}} \PP_n\big[\dist A\big]^{1/K_{d+1}}
			\leq 3\max_{A\in\{C,E,T\}} \PP_n\big[\dist A\big]^{1/K_{d+1}}.
			\ee
			Combined with Lemma \ref{refpos:lem:bound-bad-events}
			and \eqref{refpos:eq:peierls:2} we get
			\be
			\PP_n\Big[\bigcap_{i=1}^{\floor{\varphi N}}\theta_{t_i}B\Big]
			\leq 3^{\floor{\varphi N}}
			\left\{2^{d+2} \e{-(1-u)R/4}+\e{-d2^d R}+
			(C_1(1-u)n)^{-1/5}\right\}^{\floor{\varphi N}}
			\ee
			Putting this back in \eqref{refpos:eq:peierls:1}, the summand no
			longer depends on the choice of $\Gamma$ or on the subset 
			$\{q_1,..,q_{\floor{\varphi N}}\}\se\Gamma$.  For a given 
			$\Gamma$ with length $|\Gamma|=N$, the
			number of choices of subset is
			\be
			\binom{N}{\floor{\varphi N}}\leq 
			\Big(\frac {N \r e}{\floor{\varphi N}}\Big)^{\floor{\varphi N}}
			\leq (\varphi^{-1}\r e)^{\varphi N+1},
			\ee
			while the number of paths $\Gamma$ is at most
			$(2d+2)^{N-1}$.
			Combining these facts, we
			get 
			\be\begin{split}
				\PP_n[O \leftrightarrow X] &\leq
				\sum_{N\ge L/\Delta}
				C_2(d)^N \left\{2^{d+2} \e{-(1-u)R/4}+\e{-d2^d R}+
				(C_1(1-u)n)^{-1/5}\right\}^{\floor{\varphi N}}\\
				&\le 
				\left(C_3(d) \left\{2^{d+2} \e{-(1-u)R/4}+\e{-d2^d R}+
				\e{-\frac15\log(C_1(1-u)n)}\right\}
				\right)^{L/\Delta}.
			\end{split}
			\ee
			The last inequality is valid  if we set firstly $R_0$ large enough 
			(recall $R>R_0$), and then $n$ large enough (these will depend on
			$d$), so that the sum in the line before is convergent.
			Recalling that $\Delta=3d+R/n$,
			the result follows on taking $R$ large enough and then
			$n$ large enough.
		\end{proof}

		\subsection{Existence of a bad path (Lemma
			\ref{refpos:lem:SAW-exists})}
		\label{subsec:refpos:part-2}

		We describe an algorithm that takes a loop $\gamma$ connecting
		$(\ul{0},0)$ and $(\ul{x},t)$ and produces a self-avoiding walk
		$\gamma'$ in $\bbm t$ from the small cube containing $(\ul{0},0)$ to the small cube
		containing $(\ul{x},t)$, which naturally gives a path $\Gamma$ in
		$\TT$.  Essentially, $\gamma'$ is the cubes of $\bbm t$ that $\gamma$
		visits with loop-erasure applied;  we describe its construction in
		detail to keep track of how often it encounters bad cubes.
		
		First we need to extend the notions of \emph{crowded} and
		\emph{empty}, defined for big cubes, to also small cubes.
		We say that a small cube $s\in\bbm t$
		is \emph{crowded} if there are  two adjacent edges in $s$
		each containing at least one link in $s$.  Similarly, we say that
		a union of two small cubes
		$s_{\ul i,j}\cup s_{\ul i,j+1} $  (one `on top of' the other)
		is {crowded} if there are  two adjacent edges in 
		$s_{\ul i,j}\cup s_{\ul i,j+1} $ 
		each containing at least one link in  
		$s_{\ul i,j}\cup s_{\ul i,j+1} $.
		We say that a small cube $s\in\bbm t$ is \emph{empty} if it contains no
		links.   
		
		In the algorithm below, we start from $(\ul 0,0)$ and follow
		the loop $\gamma$ in an arbitrary but fixed direction until it visits
		$(\ul x,t)$ (thus we only explore one `leg' of $\gamma$).  In doing so,
		when $\gamma$ enters or exits a cube $s$ we say that it does so 
		\emph{spatially} if it traverses a link, otherwise \emph{temporally}.
		We refer to the two temporal directions of travel as 
		\emph{positive and negative time directions}.
		The following algorithm  produces $\gamma'=(s_1,\dots,s_m)$
		where $(\ul 0,0)\in s_1$ and $(\ul{x},t) \in s_m$, and at each step we
		keep track both of the sequence of cubes $s_1,\dotsc,s_{k-1}$ added so
		far, as well as the current location on our exploration of $\gamma$.
		
		\begin{enumerate}[leftmargin=*]
			\item Label the first cube of $\bbm t$ that $\gamma$ enters $s_1$,
			and follow $\gamma$ until it leaves $s_1$ for the last time. 
			\item \label{item algo reset}
			Assume that we have defined $s_1,\dots,s_{k-1}$ and that we have 
			followed $\gamma$ until it leaves
			$s_{k-1}$ for the last time
			(this is the loop erasure).  
			We may assume, by induction, that $s_1,\dots,s_{k-1}$ have been
			defined so that $\gamma$ does not visit 
			$s_1\cup\dotsb\cup s_{k-1}$ again. 
			We define $s_k$ as the cube which $\gamma$ enters when it
			leaves $s_{k-1}$ for the last time.
		\end{enumerate}
		The following steps will determine whether we 
		also define further elements
		$s_{k+1},s_{k+2}$ of the sequence before returning to step 
		\eqref{item algo reset}.
		\begin{enumerate}[resume,leftmargin=*]
			\item If $(\ul{x},t)\in  s_k$ we set $m=k$ and  stop the algorithm. 
			\item \label{item algo easy}
			Otherwise,  if $s_k$ is  crowded or empty we 
			follow $\gamma$ until it leaves $s_k$ for the last time,
			and then return to step \eqref{item algo reset}.
			\begin{figure}[tbh]
				\begin{center}
					\includegraphics[]{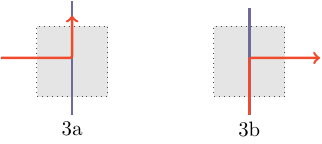}
				\end{center}
				\caption{The two possibilities for the first visit of
					$\gamma$ to a type 3 cube $s_k$ upon leaving $s_{k-1}$ for
					the last time ($\gamma$ drawn red).
				}
				\label{algorithm-3}
			\end{figure}
			
			\item If $s_k$ is neither crowded nor empty 
			we say that $s_k$ is \emph{type 3}.
			As $\gamma$ leaves
			$s_{k-1}$ for the last time and enters $s_k$,
			there are two possibilities (see Figure \ref{algorithm-3}):
			\begin{enumerate}
				\item \textit{type 3a}: $\gamma$ enters $s_k$ spatially and exits temporally,
				\item \textit{type 3b}: $\gamma$ enters $s_k$ temporally and exits spatially.
			\end{enumerate}
			Indeed, if $\gamma$ were to both enter and exit spatially then, since 
			$\gamma$ does not return to $s_{k-1}$, it would follow that 
			$s_k$ would have to be crowded.  Similarly, 
			if $\gamma$ were to both enter and exit temporally, 
			$s_k$ would have to be empty.
			
			Before proceeding to the next step of the algorithm, note that in this
			case ($s_k$ type 3) $\gamma$ cannot re-enter $s_k$.
			This is because $s_k$ is not crowded and $\gamma$ does not
			visit $s_{k-1}$ again.
			
			\item Consider the case when  $s_k$ is type 3a,
			and for simplicity 
			assume that $\gamma$ exits $s_k$  in the positive time direction
			(the case of  negative direction works similarly).  
			Write $\tilde s_{k+1}$ for the cube above $s_k$ in the time
			direction (which $\gamma$ enters after $s_k$), and
			$\tilde s_{k+2}$ for the cube above $s_{k-1}$.   
			Then one of the following occurs (see Figure \ref{fig:algorithm-3a}):
			\begin{enumerate}
				\item  The union $s_k\cup \tilde s_{k+1}$ is crowded
				or  $\tilde s_{k+1}$ is empty.  In this case,
				set  $s_{k+1}=\tilde s_{k+1}$, follow $\gamma$ until
				it leaves $s_{k+1}$ for the last time, and return to 
				step \eqref{item algo reset}. 
				\item  The above does not occur, and there is either 
				a cross
				between $\tilde s_{k+1}$ and $\tilde s_{k+2}$, or
				the cube $\tilde s_{k+2}$ is  crowded.  In either case, $\gamma$
				enters $\tilde s_{k+2}$ from $\tilde s_{k+1}$.
				We  follow $\gamma$ until it leaves the
				\emph{union} $\tilde s_{k+1}\cup \tilde s_{k+2}$ for the last
				time, set $s_{k+2}$ to be the last of the two cubes visited and
				$s_{k+1}$ to be the other one. 
				Return to step  \eqref{item algo reset}.
			\end{enumerate}
			The above are indeed the only possibilities, because the only
			remaining option is that  $\gamma$ enters $\tilde s_{k+2}$
			from $\tilde s_{k+1}$ via a double bar and returns to $s_{k-1}$ from
			above;  but by assumption, $\gamma$ does not return to $s_{k-1}$, so
			this cannot happen.
			
			
			\begin{figure}[th]
				\begin{center}
					\includegraphics[]{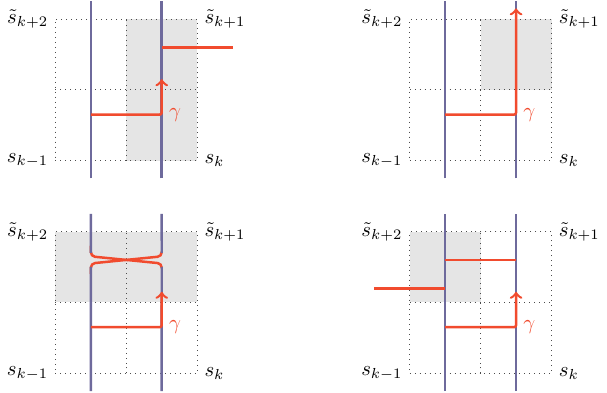}
				\end{center}
				\caption{The possibilities for case 3a.
					Top left: $s_k\cup\tilde s_{k+1}$ is crowded.
					Top right: $\tilde s_{k+1}$ is empty.
					Bottom left: $\tilde s_{k+1}\cup\tilde s_{k+2}$ has a cross. 
					Bottom right: $\tilde s_{k+2}$ is crowded. 
				}
				\label{fig:algorithm-3a}
			\end{figure}
			
			\item  \label{item algo 3b}
			Now consider the case when  $s_k$ is type 3b. 
			For simplicity, assume that
			$\gamma$ enters $s_k$ in the positive time
			direction (negative is similar).  Let $\tilde s_{k+1}$ be the cube
			$\gamma$ enters (spatially) when it exits $s_k$ and let $e$ be
			the edge connecting $s_k$ and $\tilde s_{k+1}$.
			Write  $\tilde s_{k+2}$ for the cube below 
			$\tilde s_{k+1}$  (connected to $s_{k-1}$ by $e$).  
			Then one of the following occurs (see Figure \ref{fig:algorithm-3b}):
			\begin{figure}[th]
				\begin{center}
					\includegraphics[]{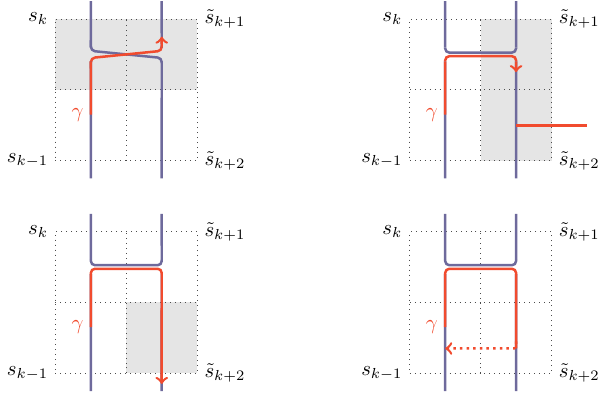}
				\end{center}
				\caption{The possibilities for case 3b.
					Top left: $s_k\cup\tilde s_{k+1}$ has a cross.
					Top right: $\tilde s_{k+1}\cup\tilde s_{k+2}$ is crowded.
					Bottom left: $\tilde s_{k+2}$ is empty. 
					Bottom right: the remaining case is that $\gamma$ returns to
					$s_{k-1}$, which is excluded by construction.  
				}
				\label{fig:algorithm-3b}
			\end{figure}	
			\begin{enumerate}
				\item If $\gamma$ enters $\tilde s_{k+1}$ via a cross,
				set $s_{k+1}=\tilde s_{k+1}$, follow $\gamma$ until it leaves 
				$s_{k+1}$ for the last time, and return to step \eqref{item algo reset}
				(recall that $\gamma$ does not
				return to $s_k$).
				\item If $\gamma$ enters $\tilde s_{k+1}$ via a double bar
				and the union $\tilde s_{k+1}\cup \tilde s_{k+2}$ is crowded,
				follow $\gamma$ until it leaves the union 
				$\tilde s_{k+1}\cup\tilde s_{k+1}$ for the last time.
				Define $s_{k+2}$ to be the last of the two cubes visited, and $s_{k+1}$
				to be the other one.  Return to step \eqref{item algo reset}.
				\item If neither of the above occurs (meaning that $\gamma$ enters 
				$\tilde s_{k+1}$   via a double bar, turns, and enters 
				$\tilde s_{k+2}$),  \emph{and}   $\tilde s_{k+2}$ is empty,
				set $s_{k+1}=\tilde s_{k+1}$, $s_{k+2}=\tilde s_{k+2}$
				and follow $\gamma$ until it leaves $s_{k+2}$ for the last time
				(note that in this case, $\gamma$ cannot return to $s_{k+1}$).
				Return to step \eqref{item algo reset}.
			\end{enumerate}
			These are  indeed the only possibilities, for if none of the above
			occur then $\gamma$ re-enters $s_{k-1}$, which is not possible (see
			Figure \ref{fig:algorithm-3b}). 
		\end{enumerate}
		This concludes the definition of 
		$\gamma'=(s_1,s_2,\dotsc,s_m)$. 
		
		Note that, by the algorithm,
		if we examine successive cubes 
		then we uncover a bad event after at most 3 cubes at a time.
		More precisely, there
		is a sequence of integers
		$k_0=1,k_1,k_2,\dotsc$ satisfying $1\leq k_i\leq 3$ for all $i\geq1$ such
		that the following holds.  With $\ell_i=k_0+k_1+\dotsb+k_i$, the $k_i$
		may be chosen so that, for all $i\geq 0$,
		$s_{\ell_i+1},\dotsc,s_{\ell_i+k_{i+1}}$ belong to a single big cube
		$S$ of some $\bb T_j$, and this cube is bad (according to our earlier
		definition).   Indeed, $k_i=1$ corresponds to the case when
		$s_{\ell_i+1}$
		itself is crowded or empty (step \eqref{item algo easy} of the
		algorithm) while $k_i\in\{2,3\}$  corresponds to the case when
		$s_{\ell_i+1}$ is type 3.
		
		
		\begin{proof}[Proof of Lemma \ref{refpos:lem:SAW-exists}]
			For each $i\in\{1,\dots,2^{d+1}\}$, define $\Gamma_i$ to be the set
			of (big) cubes in $\TT_i$ which contain as a subset at least one (small)
			cube $s_1,s_2,\dotsc,s_m$ 
			of $\gamma'$. 
			We claim that in $\bigcup_{i=1}^{2^{d+1}}\Gamma_i$, a proportion
			at least $\varphi=(2(d+1)^2+1)^{-1}$ of the cubes are bad. 
			Indeed, if $S\in \bigcup_{i=1}^{2^{d+1}}\Gamma_i$ 
			then $S$ contains at least
			one of the cubes $s_1,s_2,\dotsc,s_m$, and this cube belongs
			to one of the sequences $s_{\ell_i+1},\dotsc,s_{\ell_i+k_{i+1}}$
			considered above.   By the way the
			algorithm is defined, we may translate $S$  by
			$\pm$ one unit in at most two 
			coordinates to obtain another cube 
			$S'\in \bigcup_{i=1}^{2^{d+1}}\Gamma_i$ which
			contains the entire sequence $s_{\ell_i+1},\dotsc,s_{\ell_i+k_{i+1}}$
			and which is therefore bad.
			Thus, for each $S\in \bigcup_{i=1}^{2^{d+1}}\Gamma_i$, one out of at
			most 
			\be
			1+2(d+1)+4\binom{d+1}{2}=2(d+1)^2+1=\varphi^{-1}
			\ee
			of its translates is bad,
			meaning that  a proportion at least $\varphi$ of the cubes in 
			$\bigcup_{i=1}^{2^{d+1}}\Gamma_i$  are bad, as claimed.
			Hence there is some $\Gamma_i$ which has a proportion
			at least $\varphi$ of its cubes bad.
		\end{proof}

		
		

		\subsection{Bounds on the probabilities of bad events
			(Lemma \ref{refpos:lem:bound-bad-events})}
		\label{subsec:refpos:part-3}

		Recall that we want to bound
		$\PP_n[\dist E]$,
		$\PP_n[\dist C]$
		and $\PP_n[\dist T]$, the probabilities of the  distributed empty,
		crowded and transposition events.  
		Our strategy relies on counting loops by exploring configurations
		$\om$ from bottom to top (`time' 0 to $\beta$).
		To carry this out, we count loops using a fixed boundary
		condition rather than periodic.  The error from changing the
		boundary conditions is small if $\beta$ is large.
		
		To be more precise, write $K'=\prod_{r=1}^{d}4k_r$ for the number of
		vertices of the torus $\Lambda(\ul k)$
		and $\ell^\p(\om)$ for the number of loops in the configuration $\om$
		counting according to the periodic boundary condition.
		Next, fix any two pairings $\xi$ and $\xi'$ of the vertices of
		$\Lambda(\ul k)$, and let $\ell^{\xi,\xi'}(\om)$ be the number
		of loops of $\om$ counted by imposing
		$\xi,\xi'$ as the boundary conditions at time 0 and $\beta$,
		respectively.
		The number of loops which pass 
		through  times $0$ and $\beta$ is at least 1
		under the periodic boundary condition, and
		at most $\frac12K'+\frac12K'=K'$ under the modified boundary
		condition, so we have
		\be\label{eq loops bc 1}
		\ell^\p(\om)\leq \ell^{\xi,\xi'}(\om)+K'-1.
		\ee
		Now, consider the loops as defined using the boundary conditions
		$\xi,\xi'$.  At most $\frac12K'$ of these reach temporal level
		$\beta$, and each loop
		which does \emph{not} reach this level necessarily passes
		through a  unique highest  double bar
		(i.e.\ with maximal temporal coordinate $<\beta$ within that loop).
		Letting $L(\om)=L(\om,\xi)$ denote the number of such highest double
		bars, it follows that $\ell^{\xi,\xi'}(\om)\leq L(\om)+\frac12 K'$.
		We note that $L(\om)$  depends on $\xi$ but not on $\xi'$;  since
		$\xi$ will henceforth be arbitrary but fixed, we will often omit it
		from the notation.   
		Combined with \eqref{eq loops bc 1}, it follows that
		$\ell^\p(\om)\leq L(\om)+\tfrac32 K'$
		and therefore, for any event $A$
		\be\label{eq:crowded-boundary-conditions}
		\begin{split}
			\PP_n[A] 
			&\le 
			\frac{n^{\frac32 K'}}{Z_n}  \EE_1[n^{L(\om)} \one_{A}].
		\end{split}
		\ee
		We will argue that if $A$ is any of the events
		$\dist E$, $\dist C$ or $\dist T$ then $L(\om)$ is an order of
		magnitude smaller than $\ell^\p(\om)$, which appears in the
		denominator 
		of \eqref{eq:crowded-boundary-conditions}.
		To this end, note that we have the lower bound
		\be\label{eq pf lb}
		Z_n\ge \exp[(\tfrac{n}{2}(1-u)-d)\beta K'],
		\ee
		which can be obtained by using
		$Z_n\geq \EE_1[n^{\ell^\p(\om)}\one_A]$
		with $A$ the event that $\om$ has only double-bars restricted to some
		fixed dimer cover of $\Lambda(\ul k)$.
		\\
		
		We write the crowded event $C$ as
		$C=\bigcup_{e\sim e'\in S_{\ul 0,0}} C_{e,e'}$
		where for adjacent edges $e,e'$ in $S_{\ul{0},0}$, we write
		$C_{e,e'}$ for the event that both carry a link in
		$S_{\ul{0},0}$.
		Note that by subadditivity of the chessboard estimate,
		\cite[Lemma~5.9]{biskup-RP}, 
		we have 
		\be\label{eq:refpos:crowded-subadditivity}
		\begin{split}
			\PP_n[\dist C]^{R/\beta nK}
			&\le
			\sum_{e,e'} \PP_n[\dist C_{e,e'}]^{R/\beta nK}.
		\end{split}
		\ee
		We say that a link \emph{closes a loop} if it is the highest double
		bar on that loop.
		Recall that $\TT$ is the set of big cubes in
		the continuous torus
		$\c T=[0,4k_1]\times\dotsb\times[0,4k_d]\times[0,\beta]$.
		
		\begin{lemma}\label{refpos:lem:switches-exist}
			Let $F$ be the event that $\om$ contains at least 
			$m_0:=n\beta K/4R=|\TT|/4$ links which
			\emph{do not} close a loop.  Then for $A$ either of 
			$C_{e,e'}$ or $T$, we have that $\dist A\se F$.
		\end{lemma}
		\begin{proof}
			The lemma is clearly true for $A=T$, since on the event $\dist T$,
			there are at least $K\beta n/R$ crosses,  all of which do not close a
			loop. 
			
			Consider now the case $A=C_{e,e'}$.  A pair $(z_1,z_2)$ of links in
			$\om$, with coordinates $z_1=(e_1,t_1)$ and $z_2=(e_2,t_2)$, 
			is called a \emph{switch} if $e_1\neq e_2$ are incident
			(share a vertex), and
			$t_2>t_1$ is the \emph{minimal} temporal coordinate where an edge
			incident or equal to $e_1$ carries a link;  see  Figure
			\ref{fig:single-switch}.
			If $(z_1,z_2)$ form a switch, then both of them cannot close a loop:
			either at least one is a cross (which can never close a loop), or both
			are double bars in which case $z_2$ cannot close a loop.
			We claim that on the event $\dist C_{e,e'}$, there are at least 
			$m_0$ disjoint switches, hence at least $m_0$
			links which do not close a loop.
			
			\begin{figure}[th]
				\centering
				\includegraphics{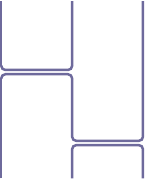}
				\caption{Two links forming a switch, illustrated in the case when both
					are double bars.  The higher link cannot close a loop.}
				\label{fig:single-switch}
			\end{figure}

			To prove the claim, first recall that for $q\in\TT$ we denote  by 
			$\theta_q$ the composition of reflections that maps $S_{\ul 0,0}$ to
			$S_q$.  We write $\TT'$ for the projection of $\TT$ on the `spatial
			dimensions' $[0,4k_1]\times\dotsb\times[0,4k_d]$.
			Thus $\TT'$ consists of  $K=\prod_{r=1}^d 2k_r$
			boxes of the form
			$[i_1-\frac12,i_1+\frac32]\times\dotsb\times [i_d-\frac12,i_d+\frac32]$
			for even $0\leq i_r\leq 4k_r-1$.
			Now fix two adjacent edges $e\neq e'$ in $S_{\ul0,0}$
			and consider the set of edges
			$\dist(e,e'):=\bigcup_{q\in\TT'}\theta_q (e\cup e')$, i.e.\ the reflections of the
			two fixed edges $e,e'$.
			We first show that this set has at least $K/4$ connected components. 
			
			Indeed, there are 3 cases,
			illustrated in Figures
			\ref{refpos:fig:reflections-a} and \ref{refpos:fig:reflections-bc}: 
			(a) both $e,e'$ lie on the boundary of
			$S_{\ul{0},0}$, (b) exactly one of $e,e'$ (say $e'$) lies on the
			boundary of $S_{\ul{0},0}$, and (c) both $e,e'$
			lie in the interior of $S_{\ul{0},0}$.
			In case (a), each connected component of
			$\dist(e,e')$ has 4 edges and intersects 4 unique boxes, thus there
			are exactly $K/4$ components.
			In case (b) each connected component has size 3,
			and is of the form
			$\theta_{q_1} e \cup (\theta_{q_1} e'=\theta_{q_2} e') \cup
			\theta_{q_2} e$ for some $q_1,q_2$ adjacent in $\TT'$,
			intersecting 2 unique boxes each.  Thus there are $K/2$ components.
			In case (c), each connected component is of size 2, and is
			of the form $\theta_q e\cup \theta_q e'$ for some $q\in\TT'$, in
			particular contained in a unique box.  Thus there are $K$ components. 
			
			\begin{figure}[th]
				\begin{center}
					\includegraphics[scale=1]{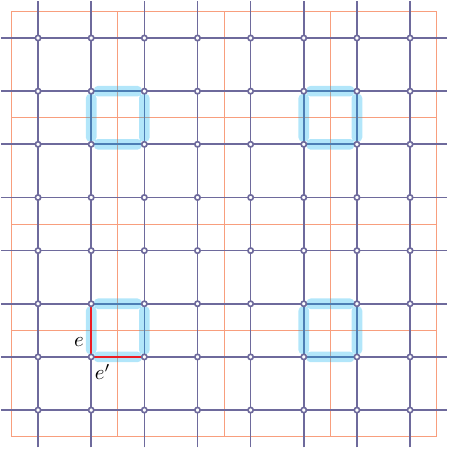}
				\end{center}
				\caption{The event $\dist C_{e,e'}$ in the case (a),
					meaning $e,e'$ are on the boundary of $S_{\ul{0},0}$. 
					The figure shows $\Lambda(\ul k)$ in the case $d=2$,
					with light orange outlining the boxes of $\TT'$.
					The components of $\dist(e,e')$ are highlighted with turquoise
					colour.  For $d=2$ as depicted here, the components are always at
					least 3 edges apart, but for $d\geq3$ it is possible for components to
					be only 1 edge apart.
				}
				\label{refpos:fig:reflections-a}
			\end{figure}
			
			\begin{figure}[th]
				\begin{center}
						\includegraphics[scale=.85]{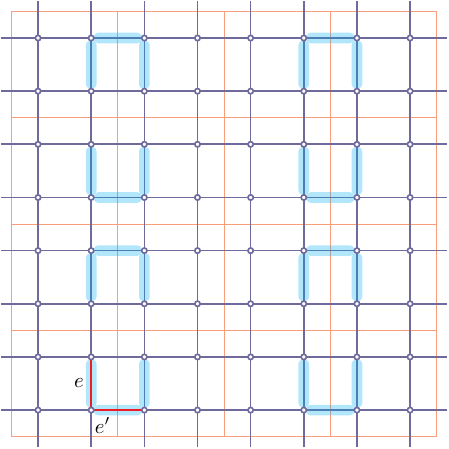}\qquad
						\includegraphics[scale=.85]{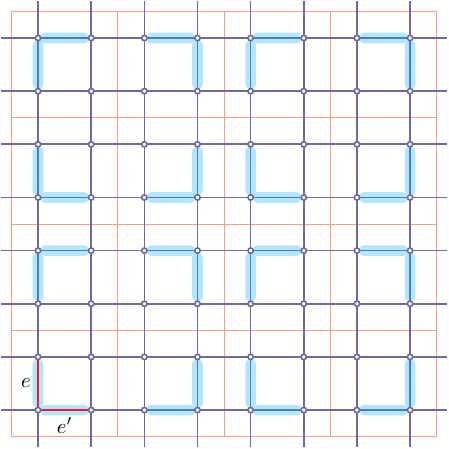}
				\end{center}
				\caption{The event $\dist C_{e,e'}$, in the cases (b)
					and (c), i.e.\ one or both edges inside $S_{\ul{0},0}$.
				}
				\label{refpos:fig:reflections-bc}
			\end{figure}
			
			The set of big boxes $\TT$ consists of $n\beta/R$ \emph{slabs},
			i.e.\ sets of the form $\TT'\times[j\frac Rn,(j+1)\frac Rn]$
			with $0\leq j\leq \beta n/R-1$.
			For $\om\in\dist C_{e,e'}$, all the edges in $\dist (e,e')$
			carry a link in each slab.  Using this, 
			we can associate to each connected component of
			$\dist(e,e')$ one switch as follows.
			First let  $\tilde z_1=(e_1,\tilde t_1)$
			be the lowest link on an
			edge of the component within the slab.  Letting $t$ increase from
			$\tilde t_1$, we eventually encounter a link $z_2$ in the same slab,
			on some edge
			$e_2\neq e_1$ incident to $e_1$.  Letting $z_1$ be the last link on
			$e_1$ encountered before $z_2$, then $(z_1,z_2)$ is a switch.
			Note that $e_1$ lies in the connected component, while $e_2$ need
			not.  Since at most two switches can share the same upper link $z_2$,
			the number of disjoint switches per slab is at least half the number
			of connected components.  Thus, on $\dist C_{e,e'}$,
			in cases (b) and (c) where the number of components is at least $K/2$,
			there are at
			least $K/4$ disjoint switches per slab, for a total of at least
			$n\beta K/4R=m_0$ disjoint switches altogether.
			
			In case  (a) the same argument gives
			at least $K/8$ disjoint switches per slab, but we can improve this to
			$K/4$ by associating \emph{two} switches to each component rather than one.
			Indeed, in case (a) each component is a cycle consisting of four
			edges $e_1,e_2,e_3,e_4$, and on $\dist C_{e,e'}$ each of these
			four edges has a link in
			each slab.  By considering all the possible relative orders of these
			links, one can check that among $e_1,e_2,e_3,e_4$ there are two
			non-incident edges $e_r$ and $e_s$ (i.e.\ not sharing a vertex) and
			two switches $(z_1,z_2)$ and $(z_1',z_2')$ such that the lower links
			$z_1$ and $z_1'$  lie on $e_r$ and $e_s$, respectively.
			Thus each slab contains $2\cdot K/4$ switches whose lower links are
			all on non-incident edges.  
			The higher links may be shared by such switches (in the same or
			different components), but since a link is shared by at most two
			different switches, we still have at least $K/4$ disjoint switches in
			each slab,  for a total of at least
			$n\beta K/4R=m_0$ disjoint switches altogether. This completes the proof of Lemma \ref{refpos:lem:switches-exist}.
		\end{proof}

		\begin{proof}[Proof of Lemma \ref{refpos:lem:bound-bad-events}]
			We start with the events $T$ and $C_{e,e'}$.
			Let $W$ be the event that the total number of links is at most 
			$M:=\r e^2dn\beta K'$ and recall that $dK'$ is the number 
			of edges in   $\Lambda (\ul k)$.
			By \cite[Theorem~1.1]{georgii-kuneth}, the process of links is
			stochastically dominated by a Poisson process of intensity $n$, so the
			number of links is stochastically dominated by a Poisson random
			variable with mean $dn\beta K'$.  Thus, by large deviations
			(if $X$ is Poisson distributed with mean $\rho$  then
			$\PP(X> K\rho)\leq \e{-\rho K\log(K/\r e)}$)
			we have that
			$\PP_n[W]\ge 1- \e{-ndK'\beta}$.
			Thus, using also \eqref{eq:crowded-boundary-conditions}
			and Lemma \ref{refpos:lem:switches-exist}
			we have for $A$ either of $T$ or $C_{e,e'}$ that
			\be\label{eq DA}
			\PP_n[\dist A] - \e{-n dK' \beta}  \le \PP_n[\dist A\cap W]
			\le \PP_n[F\cap W]
			\leq \frac{n^{\frac32 K'}}{Z_n}\EE_1[n^{L(\om)}\one_{F\cap W}],
			\ee	
			where we recall that $F$ is the event that $\om$ contains at least 
			$m_0=n\beta K/4R$ links that do not close a loop.
			Let $B_m$ denote the event that $\om$ has exactly $m$ links.  Then
			\be\label{eq LFW}
			\EE_1[n^{L(\om)}\one_{F\cap W}]\leq
			\sum_{m=m_0}^M \EE_1[n^{L(\om)}\one_F\mid B_m]\PP_1(B_m).
			\ee
			We focus now on the expectation $\EE_1[n^{L(\om)}\one_F\mid B_m]$
			with $m$ fixed.  Here the
			links are $m$ independent random variables, with uniformly chosen
			locations and types $\cross$ or $\dbar$ chosen with probabilities $u$
			and $1-u$.  Let us denote these links by $z_1,z_2,\dotsc,z_m$
			where $z_i=(e_i,t_i,\tau_i)$ and where the temporal coordinates 
			$0<t_1<t_2<\dotsb<t_m<\beta$ are
			ordered increasingly.  We condition $\EE_1[\cdot\mid B_m]$ also on the
			values of $t_1,t_2,\dotsc,t_m$ and for simplicity write 
			$\EE[\cdot]$ for the conditional measure.  
			The remaining randomness is in the uniformly chosen edges $e_i$ 
			as well as the
			types $\tau_i\in\{\cross,\dbar\}$, which are all independent.  
			
			Recall that we have fixed a pairing $\xi=:\xi_0$ 
			of the vertices of $\Lambda(\ul k)$
			which plays the role of a boundary condition	at
			time $0$.  
			Assume that we have revealed the links
			$z_1,\dots,z_{j-1}$.   This information, along with
			$\xi_0$, produces  a pairing $\xi_{j-1}$ 
			of the  vertices obtained by following the loop segments
			strictly below $z_j$, see Figure \ref{fig:pairings}.
			We call a pair $xy\in\xi_{j-1}$ \emph{minimal} if $x$ and $y$ are
			adjacent in $\Lambda(\ul k)$.
			As we place the next link $z_j$, it closes a loop 
			if and only if: 
			(i) the edge $e_j=xy$ is a minimal pair of
			$\xi_{j-1}$, and (ii) its type $\tau_j=\dbar$.
			
			\begin{figure}[th]
				\begin{center}
					\includegraphics[]{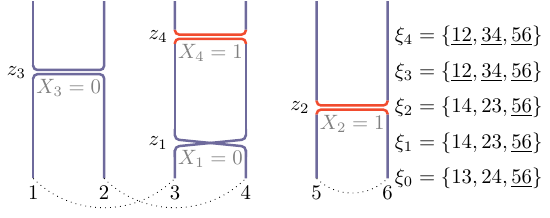}
				\end{center}
				\caption{Pairings produced by revealing
					successive links $z_j$, with minimal pairs indicated by underlining. 
					The initial pairing $\xi_0$ is indicated
					with dotted lines.
					Later pairings $\xi_j$ can be determined by, for any vertex 
					$x$, starting at $(x,t_j^+)$ and moving downwards,
					following   the loop and applying the boundary condition $\xi_0$,
					until reaching $(y,t_j^+)$ for some other vertex $y$.
					Repeating this reveals all the pairs $xy\in\xi_j$.
					The two links $z_2$
					and $z_4$ that close loops are drawn orange.
				}
				\label{fig:pairings}
			\end{figure}

			Consider the probability that $z_j$ closes a loop,
			given  $z_1,\dotsc,z_{j-1}$.
			The number of pairs in a pairing of the vertices of 
			$\Lambda(\ul k)$ is exactly 
			$\frac12 |\Lambda(\ul k)|=\frac1{2d}|\c E(\Lambda(\ul k))|$, 
			hence 
			the probability that $e_j$ is a minimal pair of  $\xi_{j-1}$
			is at most $\frac1{2d}$.  Thus the
			conditional  probability that $z_j$ closes a loop is at most 
			$\frac1{2d}(1-u)$.
			Let us write $X_j$ for the indicator that $z_j$ closes a loop,
			so $L(\om)=\sum_{j=1}^m X_j$.
			The argument above gives us that $(X_1,\dotsc,X_m)$
			is stochastically dominated by a sequence 
			$(Y_1,\dotsc,Y_m)$ of i.i.d.\ Bernoulli random variables with 
			probabilities $\frac1{2d}(1-u)$.
			(We can make this explicit by coupling the $X_j$ and $Y_j$ as
			follows.  Let $U_1,\dotsc, U_m$ be i.i.d.\ uniform on $[0,1]$.  Fix a
			partition of $[0,1]$ into $2 |\c E(\Lambda(\ul k))|$ intervals, 
			of lengths $u/|\c E(\Lambda(\ul k))|$ and $(1-u)/|\c E(\Lambda(\ul k))|$
			and each
			corresponding to an edge $e$ and a choice $\dbar$ or $\cross$.  For
			each pairing $\xi$, let $G(\xi)$ be the union of those intervals that
			correspond to minimal pairs of $\xi$ and mark $\dbar$, and let 
			$G'(\xi)\supseteq G(\xi)$ be a set containing $G(\xi)$ with measure
			exactly $\frac1{2d}(1-u)$.  Setting $X_j=\one\{U_j\in G(\xi_{j-1})\}$
			and $Y_j=\one\{U_j\in G'(\xi_{j-1})\}$ gives the desired coupling.)
			
			Recall that we write $\EE[\cdot]$ for $\EE_1[\cdot]$
			conditioned on there being a total of $m$ links as well as on their
			temporal locations.
			Using the above we get
			\be\begin{split}
				\EE[n^{L(\om)}\one_F]&\leq 
				\sum_{\substack{J\se \{1,\dotsc,m\} \\ |J|=m_0}}
				\EE[n^{\sum_{j\not\in J}X_j}]
				\leq
				\sum_{\substack{J\se \{1,\dotsc,m\} \\ |J|=m_0}}
				\EE[n^{\sum_{j\not\in J}Y_j}]\\
				&=\binom{m}{m_0}
				\big( \tfrac{n}{2d}(1-u)+\tfrac{1}{2d}(2d-1+u) \big)^{m-m_0},
			\end{split}\ee
			where we used that on $F$ there is some subset $J$ of exactly $m_0$
			links which do not close a loop.
			Recalling that $m\leq M=\e{2}dn\beta K'$
			and $m_0=n\beta K/4R$ where 
			$K'=\prod_{r=1}^d 4k_r=2^d K$, we have
			\be
			\binom{m}{m_0}\leq \Big(\frac{m\r e}{m_0}\Big)^{m_0}
			\leq \big(\e3d2^{d+2}R\big)^{m_0}.
			\ee
			Putting the above back into \eqref{eq LFW}, we get
			\be\begin{split}
				\EE_1[n^{L(\om)}\one_{F\cap W}]\leq 
				\big(\tfrac{n(1-u)+(2d-1+u)}
				{\e3d^22^{d+3}R}\big)^{-m_0}
				\sum_{m=m_0}^M
				\big( \tfrac{n}{2d}(1-u)+\tfrac{1}{2d}(2d-1+u) \big)^{m}
				\PP_1(|\om|=m)
			\end{split}\ee
			Here
			\be\begin{split}\label{eq mgf}
				\sum_{m=m_0}^M&
				\big( \tfrac{n}{2d}(1-u)+\tfrac{1}{2d}(2d-1+u) \big)^{m}
				\PP_1(|\om|=m)
				\leq 
				\EE_1\big[\big( \tfrac{n}{2d}(1-u)+\tfrac{1}{2d}(2d-1+u)
				\big)^{|\om|}\big]\\
				&=	\exp\left[
				(\tfrac{n}{2}(1-u) + \tfrac{1}{2}(2d-1+u)-1)  K'\beta
				\right].
			\end{split}\ee
			From \eqref{eq DA}, with $A=T$ or $C_{e,e'}$, we get 
			\be\begin{split}
				\PP_n[\dist A]&\leq \e{-nd K'\beta}+
				\frac{n^{\frac32 K'}}{Z_n}
				\big(\tfrac{n(1-u)}{C(d,R)}\big)^{-m_0}
				\exp\left[
				(\tfrac{n}{2}(1-u) + \tfrac{1}{2}(2d-1+u)-1)  K'\beta
				\right]\\
				&\leq \e{-nd K'\beta}+
				n^{\frac32 K'}
				\big(\tfrac{n(1-u)}{C(d,R)}\big)^{-m_0}
				\exp\big[ (2d-1-\tfrac12(1-u))  K'\beta \big],
			\end{split}\ee
			where $C(d,R)={\e3d^22^{d+3}R}$ and we used the lower bound 
			\eqref{eq pf lb} on $Z_n$.
			Using the inequality $(a+b)^x\leq a^x +b^x$ for $a,b>0$ and
			$x\in[0,1]$ we get
			\be
			\PP_n[\dist A]^{R/nK\beta}\leq
			\e{-d2^d R}+
			n^{2^{d+1}R/n\beta}
			\big(\tfrac{n(1-u)}{C(d,R)}\big)^{-\frac14}
			\exp\big[(2d-2+\tfrac{u}{2})2^dR/n\big].
			\ee
			The last factor in the second term converges to 1 as $n\to\oo$ (for
			any $R$) and by letting $\beta>20\cdot 2^{d+1}R/n$, we get
			\be
			\PP_n[\dist A]^{R/nK\beta}\leq
			\e{-d2^d R}+
			C_1(d,R) (n(1-u))^{-\frac15}.
			\ee
			This finishes the proof of Lemma \ref{refpos:lem:bound-bad-events} for the transposition event $T$ (simply taking
			$A=T$).  For the crowded event $C$ we use
			\eqref{eq:refpos:crowded-subadditivity}
			which gives a factor
			$2^d{2d\choose 2}$ for the number of choices of adjacent edges $e,e'$
			in $S_{\ul{0},0}$ (choose a vertex and then choose two edges incident
			to it).\\
			
			It remains to prove the stated bound for the `empty' event $E$, which
			we recall is the event that there is a
			small cube $s_{\ul{i},j}\subset S_{\ul{0},0}$ which is
			empty (no links).  Let $E_{\ul{i},j}$ be the event that $E$ occurs with chosen
			$s_{\ul{i},j}$. From the subadditivity of the chessboard estimate, we
			have 
			\be\label{eq:refpos:empty-subadditivity}
			\begin{split}
				\PP_n[\dist E]^{R/\beta nK}
				&\le
				\sum_{\ul{i},j} \PP_n[\dist E_{\ul{i},j}]^{R/\beta nK}
				=
				2^{d+1} \PP_n[\dist E_{\ul{i},j}]^{R/\beta nK},
			\end{split}
			\ee
			where the equality is from symmetry.
			Let $V_0\se\Lambda(\ul k)$ be the set of reflections (under
			$\theta_q$ for $q\in\TT'$) of the
			vertex at the centre of $s_{\ul i,j}$, thus $V_0$ consists of a
			fraction $1/2^d$ of all vertices of $\Lambda(\ul k)$.
			On the event $\dist E_{\ul{i},j}$, there is a subset $I\subset[0,\beta]$ of
			total length $=\beta/2$ such that  for all times $t\in I$, 
			the vertices in $V_0$  have no link on any incident
			edge.    In any pairing $\xi$ of the vertices, the number of pairs with no
			endpoint in $V_0$ is at most
			$\frac12|\Lambda(\ul k)|-\frac12|V_0|=
			\frac12(1-\frac1{2^d})|\Lambda(\ul k)|$.
			Hence on the event $\dist E_{\ul{i},j}$, for those 
			links $z_j$ such that $t_j\in I$, the probability to close a loop is
			at most $\tfrac{1}{2d}(1-\tfrac{1}{2^d})(1-u)$.  For such $j$, the
			variables $X_j$ are jointly stochastically
			dominated by i.i.d.\ Bernoulli random variables of parameter
			$\tfrac{1}{2d}(1-\tfrac{1}{2^d})(1-u)$, rather than the usual
			$\tfrac{1}{2d}(1-u)$. Putting this into working similar to
			\eqref{eq LFW}--\eqref{eq mgf}, we get
			\be\label{eq:crowded-split-event-3}
			\begin{split}
				\PP_n[\dist E_{\ul{i},j}] &\le 
				\frac{n^{\frac32 K'}}{Z_n}  \sum_{m=0}^{\infty} 
				\EE_1[n^{L(\om)}\one_{\dist F_{\ul i,j}} | B_{m}] \PP_1[B_m]\\
				&\le\frac{n^{\frac32K'}}{Z_n} 	\EE_1\left[
				\left( \tfrac{n}{2d}(1-u)+\tfrac{1}{2d}(2d-1+u) 
				\right)^{\# \text{links} \notin I} \right. \\
				&\qquad\qquad\cdot
				\left.\left( \tfrac{n}{2d}(1-\tfrac{1}{2^d})(1-u)+1
				-\tfrac{1}{2d}(1-\tfrac{1}{2^d})(1-u) \right)^{\# \text{links} \in I}
				\right]\\
				&=
				\frac{n^{\frac32K'}}{Z_n}   
				\exp\left[
				\left(\tfrac{n}{2}(1-\tfrac{1}{2^{d+1}})(1-u) 
				-\tfrac{1}{2}(1-\tfrac{1}{2^{d+1}})(1-u) \right)  K'\beta
				\right].
			\end{split}
			\ee
			Using our lower bound \eqref{eq pf lb} for $Z_n$ and
			\eqref{eq:refpos:empty-subadditivity} we get  
			\be
			\begin{split}
				\PP_n[\dist E]^{R/\beta nK}
				&\le  2^{d+1}   n^{3R2^{d-1}/\beta n}  
				\exp\left[   -\tfrac{1}{4}(1-u)R + 2^d\tfrac{d}{n}
				\right].
			\end{split}
			\ee
			The second factor tends to 1 as $n\to\infty$, and the second
			term in the exponent tends to 0, so for $n$ large enough, the
			above is bounded above by $2^{d+2} \e{-(1-u)R/4}$. 
		\end{proof}

		\section{Reflection-positivity and the chessboard estimate}
		\label{subsec:refpos:chessboard}
		
		We now prove Proposition \ref{refpos:prop:chessboard}.
		We use the  standard set-up for reflection-positivity
		\cite{biskup-RP}
		with reflections through edges of $\Lambda(\ul{k})$, and additionally allow
		reflections in the `vertical' direction.
		Let $\Pi$ be the set of all hyperplanes which are each
		orthogonal to one of the coordinate axes of $\bb R^{d+1}$
		and which, if orthogonal to one of the first $d$ (spatial) coordinates,
		pass through midpoints of edges of $\Lambda(\ul k)$.
		(So $\pi\se\Pi$ where $\pi$ is the set of hyperplanes considered in 
		Proposition \ref{refpos:prop:chessboard}.)
		For $p\in\Pi$, write $\theta_p$ for the reflection in $p$. 
		Recall that $\c T=[0,4k_1]\times\dotsb\times[0,4k_d]\times[0,\beta]$
		denotes the continuous torus in $\bb R^{d+1}$.
		Write $p_L, p_R$ for the two halves of the torus $\c T$ either side of
		the hyperplane $p$, so $p_L\cap p_R=p$. 
		
		As mentioned previously, the loop measure $\PP_n$
		itself is not reflection positive, so we will define another measure 
		$\ol \PP$ which is
		reflection-positive to which $\PP_n$ is coupled.
		Here $\ol\PP$ will be a probability measure
		on pairs $(\eta,\s)$ where $\eta$ is a point process on 
		$\c E(\Lambda(\ul k))\times[0,\beta]\times\c P$
		for some finite set $\c P$, and $\s$ is a c\`adl\`ag function
		$\Lambda(\ul k)\times[0,\beta]\to\{1,\dotsc,n\}$ with finitely many
		discontinuities (note the appearance of the parameter $n$).
		The interpretation is that $\sigma$ is a space-time
		spin configuration and  $\eta$ is a set of restrictions on the values
		of $\s$ at neighbouring vertices.  We write 
		$\Omega_{\c P}\times\Sigma$ 
		for the set of such pairs $(\eta,\s)$.

		\begin{definition}
			Let $\mu$ be a probability measure on $\Om_{\c P}\times \Sigma$ 
			and let $p\in\Pi$.  We say that $\mu$ is \emph{reflection positive} 
			across the plane $p$ if for
			all  functions $f,g$ measurable with respect to the restrictions of
			$\eta$ and $\sigma$ to  $p_L$, we have that 
			\be\label{refpos:eq:refpos-defn}
			\begin{split}
				&\mu[f\cdot\theta_pf]\ge0,\\
				&\mu[f\cdot\theta_pg]=\mu[g\cdot\theta_pf].
			\end{split}  
			\ee
		\end{definition}
		
		To define the precise measure $\ol\PP$ that we work with, we follow
		Ueltschi \cite[Sections 2 and 5]{ueltschi}.
		Let $\c P$ denote the power set (set of all subsets) of 
		$\{1,\dotsc,n\}^4$.  Then $\eta$ is a set of points on 
		$\c E(\Lambda(\ul
		k))\times[0,\beta]$  labelled (marked) with some set 
		$A\in\c P$ of tuples $(a,b,c,d)\in\{1,\dots,n\}^4$.
		Let $<$ denote an arbitrary but fixed total order of the vertices
		$\Lambda(\ul k)$.
		We display elements of $\{1,\dots,n\}^4$ in the form
		$\pattern abcd$, and they designate a colouring of 
		$(\ul x,t\pm)$ and $(\ul y,t\pm)$ 
		assigned to an edge $\{\ul{x},\ul{y}\}$, where
		$\ul{x}<\ul{y}$, just before and just after time $t$.
		An element $A\in\c P$ thus gives a set of `allowed' 
		colourings $\pattern abcd$.  More precisely, given a point
		configuration $\eta\in\Omega_{\c P}$, we say that $\s\in\Sigma$ is
		\emph{compatible}  with $\eta$ if for each discontinuity point 
		$(\ul{x},t)$ of $\sigma$ there is a mark of $\eta$ at $(e,t)$
		for some $e=\{\ul x,\ul y\}\ni \ul x$, and the values of $\s$
		at $(\ul x,t\pm)$ and $(\ul y,t\pm)$ belong to the corresponding set
		$A$ with which the mark of $\eta$ is labelled.
		We write $\s\sim\eta$ if $\s$ is comptabile with $\eta$.
		
		For each $e\in \c E(\Lambda(\ul k))$
		let $\iota_e:\c P\to \RR_{\ge0}$ be a collection of 
		non-negative intensities associated with the edge $e$.  
		Let $\PP_\iota$ be the probability measure under which 
		$\eta$ is a superposition of independent Poisson
		processes, of intensities $\iota_e(A)$ for each edge $e$ and set
		$A\in\c P$.  We define the 
		probability measure $\ol\PP_\iota$ on 
		pairs $(\eta,\sigma)\in\Om_{\c P}\times\Sigma$ 
		by 
		\be
		\ol\PP_\iota[A]=\frac1{Z_\iota}\EE_\iota\Big[
		\sum_{\sigma\sim\eta} \one\{(\eta,\sigma)\in A\}\Big]
		\ee
		where $Z_\iota$ is the appropriate normalization.  Informally we may
		think of $\ol\PP_\iota$ as  the product of the Poisson law on $\eta$
		with the counting measure on $\sigma$, restricted to compatible
		pairs.

		To make a connection with our loop-measure $\PP_n$,
		consider the following intensity:
		\be
		\begin{split}
			\iota^{\ssc{1}}
			\left( \big\{ \pattern abba: 
			1\leq a\neq b\leq n
			\big\} \right)&= u, \\
			\iota^{\ssc1}
			\left( \big\{ 
			\pattern aabb: 
			1\leq a\neq b\leq n
			\big\} \right)	&= 1-u,\\  
			\iota^{\ssc1}
			\left( \big\{ 
			\pattern cccc: 1\leq c\leq n
			\big\} \right)	&= 1,\\  
			\iota^{\ssc1}
			(A)
			&= 0,
			\quad\quad \text{for all other }A\in\c P. 
		\end{split}
		\ee
		Given a sample $(\eta,\sigma)\sim\ol\PP_{\iota^{\ssc1}}$, form a
		link-configuration $\om$ by placing a cross $\cross$
		wherever $\eta$ has a mark
		$\big\{  \pattern abba:  1\leq a\neq b\leq n\big\}$,
		placing a double-bar $\dbar$ wherever $\eta$ has a mark
		$\big\{ \pattern aabb :  1\leq a\neq b\leq n \big\}$,
		and placing $\cross$ or $\dbar$ with probabilities $u$ and $1-u$
		wherever $\eta$ has a mark
		$\big\{ \pattern cccc :  1\leq c\leq n \big\}$.
		The marginal distribution on $\omega$ is then exactly $\PP_n$, since
		there are $n^{\ell(\om)}$ possible choices of the colouring $\sigma$.
		However, this representation is not reflection-positive, since
		the patterns $\pattern abba$ are not reflection-symmetric.  We
		therefore define a different intensity function where we combine 
		$\pattern abba$ with $\pattern aabb$ to get symmetric constraints.  This is
		only possible if crosses are less frequent than double-bars, 
		i.e.\ if $u\leq \frac12$.
		
		Consider the following subsets of $\{1,\dots,n\}^4$:
		\be\label{eq roman sets}
		\begin{split}
			&\texttt{I}_{a,b}  = 
			\big\{ \pattern aabb,  \pattern abba , \pattern bbaa,  \pattern baab 
			\big\},\quad \text{for } 1\leq a< b\leq n,\\
			&\texttt{II}_{a,b} =
			\big\{ \pattern aabb \big\}, \quad \text{for }  1\leq a\neq b\leq n,\\
			&\texttt{III}_c = \big\{ \pattern cccc \big\}, \quad \text{for } 1\leq c\leq n.    
		\end{split}
		\ee
		Let $\pi'\se \Pi$ be a collection of hyperplanes that
		is preserved by the reflection $\theta_p$ for any
		$p\in\pi'$.  Note that the collection $\pi$ of planes separating cubes of
		$\TT$ defined in Section \ref{subsec:refpos:part-1} is such a
		collection.  For all edges $e$ which a plane $p\in\pi'$ intersects,
		define $\iota^{\ssc2}_e=\iota^{\ssc2}$ as: 
		\be\label{eq iota2}
		\begin{array}{ll}
			\iota^{\ssc2}
			\left( \texttt{I}_{a,b} \right)=   u
			&\text{for } 1\leq a< b\leq n,\\
			\iota^{\ssc2}
			\left( \texttt{II}_{a,b} \right)=   1-2u
			& \text{for } 1\leq a\neq b\leq n, \\  
			\iota^{\ssc2}
			\left( \texttt{III}_c \right) =  1
			& \text{for } 1\leq c\leq n,\\ 
			\iota^{\ssc2}(A)=   0
			& \text{for all other }A\in\c P. 
		\end{array}
		\ee
		For edges $e$ which do not intersect a hyperplane $p\in\pi'$, let
		$\iota^{\ssc2}_e=\iota^{\ssc1}$.
		Consider the resulting measure $\ol\PP_{\iota^{\ssc2}}$.
		We augment this measure by including i.i.d.\
		random variables that allow to 
		determine the choices between $\cross$ and $\dbar$ at marks 
		$\big\{ \pattern cccc: 1\leq c\leq n\big\}$ for edges $e$ not bisected
		by planes $p\in\pi'$, i.e.\ for those edges where the intensity is
		$\iota^{\ssc1}$.

		We also need \emph{Mecke's formula},
		which provides a method for conditioning on the exact locations of
		points in a Poisson process (see e.g.\
		\cite[Theorem 4.4]{last-penrose}).
		Recall that $\PP_1$ is the law of a Poisson process on 
		$\c E(\Lambda)\times[0,\beta]\times\{\cross,\dbar\}$.
		We write $\mu$ for its intensity measure, which is a product of $u$
		times Lebesgue measure on $\c E(\Lambda)\times[0,\beta]$
		with $1-u$ times the same.
		
		\begin{lemma}[Mecke's formula]
			\label{lem mecke}
			For $f$
			any bounded measurable function of pairs of finite subsets of 
			$\c E(\Lambda)\times[0,\beta]\times\{\cross,\dbar\}$,
			\be
			\EE_1\Big[
			\sum_{\eta\se\omega, |\eta|=m} f(\eta,\omega)
			\Big]
			=\int \dd \mu^{\odot m}(\eta)\; \EE_1[f(\eta,\omega\cup\eta)]
			\ee
			where 
			\be
			\dd \mu^{\odot m}(\{x_1,\dotsc,x_m\})
			=\frac1{m!} \sum_{\pi\in S_m}
			\dd\mu^{\otimes m}(x_{\pi(1)},\dotsc,x_{\pi(m)})
			\ee
			is the symmetrized $m$-fold product measure.
		\end{lemma}
		
		\begin{proposition}\label{refpos:prop:PP-iota-refpos}
			Let $\pi'$ and $\ol\PP_{\iota^{\ssc2}}$ be as above. 
			Then $\ol\PP_{\iota^{\ssc2}}$ is reflection positive across any $p\in\pi'$.
		\end{proposition}
		\begin{proof}
			The argument is the standard one for showing that a measure is reflection
			positive:  we show that conditional on the configuration on a plane
			$p\in\pi'$, the configurations on the left and right halves of the
			torus are independent and identically distributed. 
			
			Let $p\in\pi'$. The measure $\PP_{\iota^{\ssc2}}$ is
			reflection symmetric in $p$ ($\theta_p\om\sim\PP_{\iota^{\ssc2}}$ if
			$\om\sim\PP_{\iota^{\ssc2}}$), so the second condition in
			\eqref{refpos:eq:refpos-defn} holds 
			(here we require the
			condition that $\pi'$ is preserved in any reflection $\theta_p$,
			$p\in\pi'$). 
			
			For the first condition, assume first that $p\in\pi'$ is orthogonal to
			a spatial coordinate.  For $(\eta,\s)\in\Om_{\c P}\times\Sigma$, 
			write $\eta_p$ for the configuration $\eta$ restricted to edges which
			intersect $p$, and $\eta_R, \eta_L$ for the configurations
			restricted to the right and left sides of the plane $p$ (both including
			the plane itself), and similarly write $\sigma_R, \sigma_L$ for the
			restrictions of $\s$ to the left and right sides.  
			We write $\s_L\sim\eta_L$
			and $\s_R\sim\eta_R$ for compatibility of the restricted
			configurations defined in the natural way.
			We have
			\be
			\begin{split}
				Z_{\iota^{\ssc2}}\ol\PP_{\iota^{\ssc2}}[f\cdot\theta_p f] &=
				\sum_{m=0}^\infty\EE_{\iota^{\ssc2}}
				\Big[
				\sum_{\substack{\tau\se\eta \\ |\tau|=m}}
				\Ind{\eta_p=\tau}
				\sum_{\s\sim\eta}
				f(\eta_L,\sigma_L)\cdot\theta_p f(\eta_R,\sigma_R)\Big] 
				\\
				&=
				\sum_{m=0}^\infty
				\int \dd \mu^{\odot m}(\tau) 
				\EE_{\iota^{\ssc2}}
				\Big[
				\Ind{(\eta\cup\tau)_p=\tau}
				\sum_{\s\sim\eta\cup\tau}
				f((\eta\cup\tau)_L,\sigma_L)
				\cdot\theta_p f((\eta\cup\tau)_R,\sigma_R) 
				\Big]\\
				&=
				\sum_{m=0}^\infty
				\int \dd \mu^{\odot m}(\tau) 
				\EE_{\iota^{\ssc2}}
				\Big[
				\Ind{(\eta\cup\tau)_p=\tau}
				\sum_{\sigma_R \sim \eta_R\cup\tau}
				f((\eta\cup\tau)_L,\sigma_L)
				\Big]\\
				&\hspace{2cm}
				\EE_{\iota^{\ssc2}}
				\Big[
				\Ind{\tau=(\eta\cup\tau)_p}
				\sum_{\sigma_R \sim \eta_R\cup\tau}
				\theta_pf((\eta\cup\tau)_R,\sigma_R)
				\Big]\\
				&=
				\sum_{m=0}^\infty
				\int \dd \mu^{\odot m}(\tau) 
				\EE_{\iota^{\ssc2}}
				\Big[
				\Ind{\tau=(\eta\cup\tau)_p}
				\sum_{\sigma_L \sim \eta_L\cup\tau}
				f((\eta\cup\tau)_L,\sigma_L)
				\Big]^2\\
				&\ge0.
			\end{split}
			\ee
			We used  Mecke's formula Lemma 
			\ref{lem mecke} to condition on the marks $\eta_p$
			(here $\mu$ is the appropriate intensity measure).
			In the third line we used that, given the configuration on $p$, the
			compatibility condition for $\sigma$ factorizes into separate
			conditions on the left and right sides. 
			In the fourth line we used the symmetry of $\PP_{\iota^{\ssc2}}$
			as well as of the boundary condition imposed by $\tau$;
			this final property is where we need the sets 
			\eqref{eq roman sets} to be symmetric under reflecting across an
			edge. 
			In the calculation above we did not explicitly mention the i.i.d.\
			random variables that allow to 
			determine the choices between $\cross$ and $\dbar$ at marks 
			$\big\{ \pattern cccc: 1\leq c\leq n\big\}$ for edges $e$ not bisected
			by planes $p\in\pi'$.  However, since they are i.i.d.\ and are not
			affected by the conditioning on $\eta_p$, the conclusion remains
			valid.

			For planes $p\in\pi'$ which are orthogonal to the temporal axis the
			argument is the same,  except that the restriction to $p$ applies
			to $\sigma$ rather than $\eta$, and is almost surely a symmetric
			condition (we do not need to use
			Mecke's formula to condition on the value of $\sigma$ on $p$). 
		\end{proof}
		
		Similarly to the case of $\ol\PP_{\iota^{\ssc1}}$,
		the measures $\ol\PP_{\iota^{\ssc2}}$ and $\PP_n$  can be coupled.
		Define the measure $\QQ_n$ on triples $(\eta,\sigma,\om)$
		as follows.
		\begin{itemize}[leftmargin=*]
			\item First sample $(\eta,\sigma)\sim\ol\PP_{\iota^{\ssc2}}$. 
			\item On edges not bisected by some $p\in\pi'$, 
			where the intensity function $\iota^{\ssc1}$ is used, let
			$\om$ be given as in the coupling with $\ol\PP_{\iota^{\ssc1}}$. 
			\item On edges $e$ bisected by some $p\in\pi'$, a link 
			of $\eta$ is marked by a set $A$ which is one of
			$\texttt{I}_{a,b}, \texttt{II}_{a,b}$ or $\texttt{III}_{c}$,
			for some $a,b$ or $c$.  Further, the choice of $\sigma$ 
			agrees with some element of $A$ at the location of the
			link.   If this element is $\pattern abba$ for some
			$a\neq b$, place a  cross $\cross$ in $\om$. 
			If the element is $\pattern aabb$ for some $a\neq b$, 
			place a double bar $\dbar$.  If the
			element is $\pattern cccc$ for some $c$, place a
			$\cross$ with probability $u$ or else a $\dbar$ with probability
			$1-u$, independently of everything else. 
		\end{itemize}

		\begin{proposition}
			If $(\eta,\sigma,\om)\sim\QQ_n$, then $\om\sim\PP_n$.
		\end{proposition}
		\begin{proof}
			To avoid introducing too much notation, we sketch the proof. 
			Taking the marginal on $\om$ amounts to, at each link of $\om$,
			summing over the possible choices for the colours of $\sigma$ at that
			link and over the possible choices $A\in\c P$ for $\eta$ compatible
			with the colours of $\sigma$.  For $e$ not bisected by planes
			$p\in\pi'$, the weights (under $\iota^{\ssc1}$) 
			for $\cross$ and $\dbar$ are explicitly $u$ and $1-u$.
			For $e$ which are bisected by planes $p\in\pi'$, the choices for $A$
			are  $\texttt{I}_{a,b}$,
			$\texttt{II}_{a,b}$ or $\texttt{III}_c$, where both $\texttt{I}$ and
			$\texttt{II}$ are possible for $\dbar$, with weights summing
			to $1-u$, while for $\cross$ only
			$\texttt{I}$ is possible and its weight is $u$.
			
			We give slightly  more detail, assuming for simplicitly that all edges
			have the intensity function $\iota^{\ssc2}$
			of \eqref{eq iota2}.
			Let $\s\sim\om$ denote the condition that $\s$ is
			constant on the loops of $\om$ and write $\ol\om$ and $\ol\eta$ for
			the supports of these configurations (i.e.\ the locations of links or
			marks).   We can write the marginal
			\be\begin{split}
				\dd \QQ_n(\om)&\propto\sum_{\s\sim\om}
				\sum_{\substack{\eta\sim\s\\\ol\eta=\ol\om}}
				\dd \ol\PP_{\iota^{\ssc2}}(\eta,\s) 
				\\
				&=\sum_{\s\sim\om}
				\sum_{\substack{\eta\sim\s\\\ol\eta=\ol\om}}
				u^{\#\texttt{I}(\eta)}(1-2u)^{\#\texttt{II}(\eta)}
				u^{\#\cross(\om\cap S(\s))}(1-u)^{\#\dbar(\om\cap S(\s))}
				\dd e^{|\om|},
			\end{split}\ee
			where $\dd e$ denotes Lebesgue measure and $S(\s)$ denotes the set of
			$(e,t)$ where $\s$ takes the same value on both endpoints.  
			Here (with $a\neq b$)
			\be
			\sum_{\substack{\eta\sim\s\\\ol\eta=\ol\om}}
			u^{\#\texttt{I}(\eta)}(1-2u)^{\#\texttt{II}(\eta)}
			=\prod_{\pattern aabb\in\s} (u+1-2u) 
			\prod_{\pattern abba\in\s} u, 
			\ee
			so we get
			\be
			\dd \QQ_n(\om)\propto\sum_{\s\sim\om}
			u^{\#\cross(\om)}(1-u)^{\#\dbar(\om)}
			\dd e^{|\om|}=n^{\ell(\om)} 
			u^{\#\cross(\om)}(1-u)^{\#\dbar(\om)}
			\dd e^{|\om|},
			\ee
			as required.
		\end{proof}
		
		Finally, we can prove the chessboard estimate.
		
		\begin{proof}[Proof of Proposition \ref{refpos:prop:chessboard}]
			Let ${\iota^{\ssc2}}$ be defined using the hyperplanes $\pi$
			passing through boundaries of cubes of $\TT$, as in 
			Section \ref{subsec:refpos:part-1}. The chessboard estimate
			applies to any measure reflection positive in the hyperplanes of a
			torus. Let $A_i$, $i=1,\dots,m$ be events measurable with respect to
			$(\eta,\s)\sim\ol\PP_{\iota^{\ssc2}}$, dependent only on the
			configuration in the cube $S_{\ul{0},0}$.
			The standard proof,  see for example \cite{biskup-RP}, then gives 
			\be
			\ol\PP_{\iota^{\ssc2}}\left[
			\bigcap_{i=1}^m \theta_{q_i} A_i
			\right]
			\le
			\prod_{i=1}^m
			\ol\PP_{\iota^{\ssc2}}\left[
			\bigcap_{q\in\TT}\theta_q A_i
			\right]^{R/nK\beta}.
			\ee
			Let
			$(\eta,\s,\om)\sim\QQ_n$. It remains to observe that an event $A$
			measurable with respect to $\om$, occurring in $S_{\ul{0},0}$ and
			independent of link type on the boundary of $S_{\ul{0},0}$, is in fact
			measurable with respect to
			$(\eta,\s)\sim\ol\PP_{\iota^{\ssc2}}$.  Indeed, under the coupling
			$\QQ_n$ the marks of $\eta$ and the links of $\om$ are in bijection
			(given by their locations), so a function of $\om$ independent of link
			type can be written as the same function of $\eta$. 
			Inside the cube $S_{\ul{0},0}$, the exact link-configuration $\omega$
			is determined since we augmented $\ol\PP_{\iota^{\ssc2}}$ to include
			the i.i.d.\ random variables that determine the choices $\cross$ and
			$\dbar$ for edges not bisected by the planes $p\in\pi$.
		\end{proof}

	\end{document}